\documentclass[11pt]{article}
\usepackage{comment,soul}
\usepackage[hidelinks]{hyperref}
\usepackage{enumitem}
\usepackage{amsmath,amssymb,epsf,cite,graphicx,subfigure}
\usepackage{amsmath,braket,tikzsymbols}
\usepackage[bbgreekl]{mathbbol}
\usepackage{comment}
\usepackage[export]{adjustbox}

\numberwithin{equation}{section}

\definecolor{airforceblue}{rgb}{0.36, 0.54, 0.66}
\newcommand{\beq}{\begin{equation}}
\newcommand{\eeq}{\end{equation}}

\begin{document}
\baselineskip=15.5pt
\pagestyle{plain}
\setcounter{page}{1}

\begin{center}
{\LARGE \bf A precision test of averaging in AdS/CFT}
\vskip 1cm

\textbf{Jordan Cotler$^{1,2,a}$ and Kristan Jensen$^{3,b}$}

\vspace{0.5cm}

{\it ${}^1$ Society of Fellows, Harvard University, Cambridge, MA 02138, USA \\}
{\it ${}^2$ Black Hole Initiative, Harvard University, Cambridge, MA 02138 USA \\}
{\it ${}^3$ Department of Physics and Astronomy, University of Victoria, Victoria, BC V8W 3P6, Canada\\}

\vspace{0.3cm}

{\tt  ${}^a$jcotler@fas.harvard.edu, ${}^b$kristanj@uvic.ca\\}

\medskip

\end{center}

\vskip1cm

\begin{center}
{\bf Abstract}
\end{center}
\hspace{.3cm} 
We reconsider the role of wormholes in the AdS/CFT correspondence. We focus on Euclidean wormholes that connect two asymptotically AdS or hyperbolic regions with $\mathbb{S}^1\times \mathbb{S}^{d-1}$ boundary. There is no solution to Einstein's equations of this sort, as the wormholes possess a modulus that runs to infinity. To find on-shell wormholes we must stabilize this modulus, which we can do by fixing the total energy on the two boundaries. Such a wormhole gives the saddle point approximation to a non-standard problem in quantum gravity, where we fix two asymptotic boundaries and constrain the common energy. Crucially the dual quantity does not factorize even when the bulk is dual to a single CFT, on account of the fixed energy constraint. From this quantity we extract a smeared version of the microcanonical spectral form factor. For a chaotic theory this quantity is self-averaging, i.e.~well-approximated by averaging over energy windows, or over coupling constants.

We go on to give a precision test involving the microcanonical spectral form factor where the two replicas have slightly different coupling constants. In chaotic theories this form factor is known to smoothly decay at a rate universally predicted in terms of one replica physics, provided that there is an average either over a window or over couplings. We compute the expected decay rate for holographic theories, and the form factor from a wormhole, and the two exactly agree for a wide range of two-derivative effective field theories in AdS. This gives a precision test of averaging in AdS/CFT.

Our results interpret a number of confusing facts about wormholes and factorization in AdS and suggest that we should regard gravitational effective field theory as a mesoscopic description, analogous to semiclassical mesoscopic descriptions of quantum chaotic systems.

\newpage

\newpage

\tableofcontents

\newpage
\section{Introduction and Summary}

In this work we primarily consider spacetimes that smoothly connect two asymptotically asymptotically Euclidean AdS.  These Euclidean wormholes have long posed a puzzle~\cite{witten1999connectedness, maldacena2004wormholes} for the AdS/CFT correspondence. In standard examples of holographic duality, string/M-theory on some (Euclidean) AdS vacuum is equated with a particular conformal field theory (CFT). The natural holographic dictionary with two boundaries equates the sum over bulk configurations with two asymptotically hyperbolic regions to a two-replica partition function of the dual theory,
\beq
	\includegraphics[width=3in, valign = c]{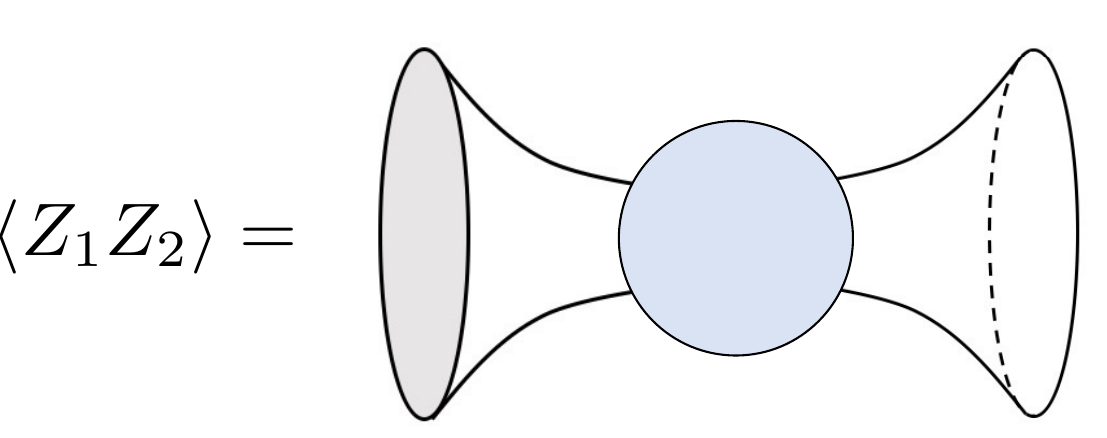}
\eeq
where the blob represents a sum over configurations that fill in the boundary data and we have allowed for a potential ensemble average denoted by the brackets. But, if the dual is a single CFT, this quantity ought to factorize, $\langle Z_1 Z_2 \rangle  \stackrel{?}{=} \langle Z_1\rangle\langle Z_2\rangle$, or in a picture,
\beq
	\includegraphics[width=5.5in, valign = c]{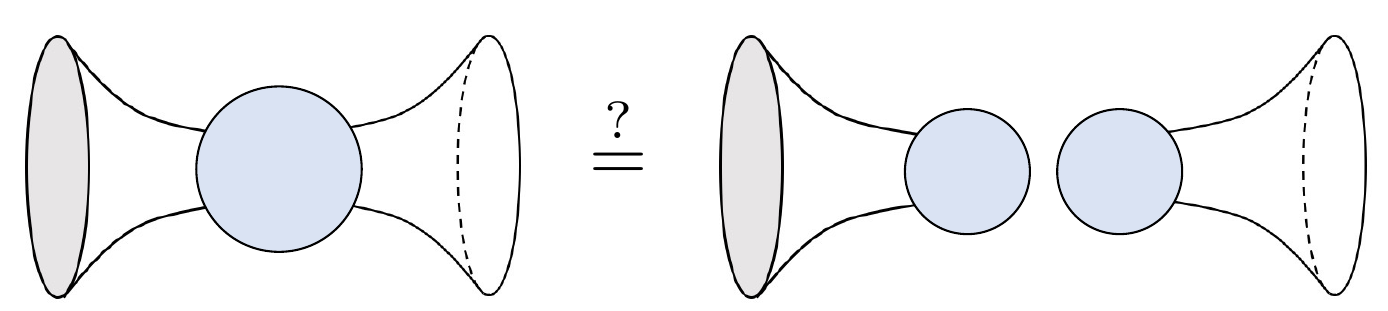}
\eeq
Wormholes pose a problem because they contribute to the connected part of $\langle Z_1 Z_2\rangle$, which if nonzero would spoil factorization of the two-boundary amplitude and, barring a modification of the holographic dictionary, indicate an ensemble average in the holographic dual. From the bulk point of view, there is no obvious reason why this connected part should vanish, but nevertheless we expect factorization in standard stringy realizations of the AdS/CFT correspondence. This tension was identified in~\cite{witten1999connectedness, maldacena2004wormholes}, which we refer to as the ``factorization paradox''.

Despite being an old puzzle, the factorization paradox has been the subject of recent attention largely because of progress in the study of simple, soluble models of quantum gravity in low spacetime dimension. Jackiw-Teitelboim (JT) gravity~\cite{Jensen:2016pah, Engelsoy:2016xyb, Maldacena:2016upp} and many of its cousins~\cite{Stanford:2019vob, maldacena2019two, cotler2019low, Maxfield:2020ale, Witten:2020wvy} are consistent theories of two-dimensional quantum gravity, defined non-perturbatively by a dual ensemble of random Hamiltonians. Wormhole amplitudes in these theories have been computed non-perturbatively in the bulk gravitational interaction, and encode the fluctuation statistics of the dual ensemble. Beyond JT gravity, some wormhole amplitudes have been computed for pure gravity in three dimensions with negative cosmological constant. Two-boundary solutions exist when the two boundaries have the same genus $g>1$, and the corresponding amplitudes are known to one-loop~\cite{Giombi:2008vd}. There is no saddle point for the two-torus amplitude, but even so we have computed it to one-loop order in our previous work~\cite{cotler2020ads}. Related amplitudes were later computed in~\cite{Chandra:2022bqq, Eberhardt:2022wlc}.  So, if pure gravity in three dimensions is a consistent theory of gravity, non-factorization of multi-boundary amplitudes implies that the dual is some suitable ensemble.\footnote{While one might wonder about an ensemble of two-dimensional conformal field theories, this possibility seems unlikely, as we already noted in~\cite{cotler2020ads}. The problem is that we do not know of a single irrational conformal field theory at large central charge, let alone a large number of them to form an ensemble.  As we noted in our previous work, it seemed more likely that the dual ensemble would be a set of (approximate) solutions to the modular bootstrap. In a recent paper, building on~\cite{Belin:2020hea, Belin:2021ryy, Anous:2021caj}, the authors of~\cite{Chandra:2022bqq} were able to match a number of wormhole saddles in three-dimensional gravity to averages over an ensemble of solutions to the ($S$-)bootstrap describing two-dimensional CFTs at large central charge with a sparse spectrum of light states.}

In fact wormholes in JT gravity do more than simply exist: they encode physics. Upon analytic continuation the two-boundary amplitude gives the spectral form factor of the dual description $\left\langle \text{tr}\!\left( e^{-(\beta+i T)H}\right)\text{tr}\!\left(e^{-(\beta-i T)H}\right)\right\rangle$, with a smooth ramp over a long timescale coming from a wormhole contribution to the amplitude, the ``double trumpet'' of~\cite{Saad:2018bqo, Saad:2019lba}. Physically, this ramp is a consequence of long-range repulsion in the spectrum of the boundary Hamiltonian, and the average over Hamiltonians leads to a smooth ramp.

The presence of a ramp is not a surprise. The energy eigenvalues in question are the black hole microstates of the model. Insofar as one expects black holes to be chaotic systems, random matrix universality predicts long-range repulsion in their spectrum of microstates, which is exactly what one finds from the wormhole amplitude. The smoothness of the ramp is a more delicate matter, which we will have more to say about later. 

Known wormhole amplitudes in pure three-dimensional gravity also encode meaningful physics. The two-torus amplitude~\cite{cotler2020ads, cotler2020ads2} is analogous to the ``double trumpet'' of JT gravity, and encodes a smooth ramp\footnote{In fact that amplitude is precisely what one would get from a random matrix theory ansatz for the connected two-point function of the density of states, augmented by Virasoro symmetry and modular invariance.} in the spectral form factor of the dual. This ramp implies repulsion in the spectrum of BTZ black hole microstates, and the breakdown of factorization suggests that we have an ensemble. If pure three-dimensional gravity is a consistent theory of quantum gravity, we infer that when we discuss the BTZ black hole (in pure gravity), we are dealing with an ensemble of chaotic Hamiltonians. As for the higher genus wormholes of~\cite{maldacena2004wormholes}, or wormholes supported by heavier operators (below the black hole threshold but whose dimension scales with the boundary central charge $c$)~\cite{Chandra:2022bqq}, these can be understood in terms of a distribution of OPE coefficients for heavy operators which is approximately Gaussian at large central charge~\cite{Belin:2020hea, Belin:2021ryy, Anous:2021caj}.

While these simple models of gravity are of interest in their own right, in this work we are interested in stringy examples of AdS/CFT like the celebrated duality between 4d $\mathcal{N}=4$ super Yang Mills theory and IIB strings on AdS$_5\times\mathbb{S}^5$. The simple models are of course quite different as they have no local bulk degrees of freedom.  That being said, one might wonder if they tell us about wormholes in stringy AdS/CFT, since these simple models emerge as effective descriptions of certain near-extremal black holes in string/M-theory. However this is not the case. The wormholes of JT gravity are off-shell configurations, and so do not uplift to wormhole saddles in higher dimensions. Similarly, in pure 3d gravity, the two torus wormholes of~\cite{cotler2020ads, cotler2020ads2} are off-shell and do not uplift; pure 3d gravity does have wormhole saddles, like the higher genus wormholes mentioned above, and these do uplift to saddles of 10/11d supergravity. However the uplifts suffer from brane nucleation instabilities. So, while results about wormholes in these theories do not directly tell us about (non-)factorization in string theoretic examples of AdS/CFT, they have brought these old questions about wormholes and factorization back to our attention.

In this work we focus mostly on Euclidean wormholes where the boundary conditions are the same ones relevant for thermal physics and black holes, in particular $\mathbb{S}^1\times\mathbb{S}^{d-1}$ or a torus, for which the bulk wormholes are always off-shell (in accordance with the Witten-Yau theorem~\cite{witten1999connectedness}).  We find two main results. First, we show how to stabilize these wormholes so that they can be the saddle point approximation to a non-standard question in AdS quantum gravity. That is we find what can be reliably computed from these wormholes in bulk effective field theory, and the ensuing implications for the factorization paradox. Our results are consistent with those of~\cite{Saad:2018bqo} regarding the ``double cone'' and the spectral form factor of holographic CFTs. Second, we use one of these reliable computations to perform a precision test whereby a wormhole describes an averaged quantity in the dual CFT. 

We now give a detailed summary our main results, compare with previous work, and discuss the implications of our findings.

\subsection{Stabilizing wormholes}

In our previous work we parameterized wormhole amplitudes where the wormhole has the topology of $\mathbb{T}^d$ times an interval, or $\mathbb{S}^1\times\mathbb{S}^{d-1}$ times an interval. The two asymptotically (Euclidean) AdS boundaries have circles of sizes $\beta_1$ and $\beta_2$ respectively. Later we will consider complex wormholes with complex $\beta_1,\beta_2$ for which $\text{Re}(\beta_1)$ and $\text{Re}(\beta_2)$ are greater than zero.  The amplitude can be written as
\beq
\label{E:wormholeAmplitude}
	Z_{\rm wormhole}(\beta_1,\beta_2) = \int_{E_0}^{\infty} dE \,e^{-(\beta_1+\beta_2)E} f(E)\left( 1 + O(G)\right)\,.
\eeq
In order to explain the ingredients in this expression, let us first discuss where it comes from. One way to arrive at it is via the method of constrained instantons~\cite{affleck1981constrained}, which we will call constrained saddles since we are studying wormholes rather than instantons in gauge theory. Foliating the space of fields according to a constraint, one employs a saddle-point approximation along the slices of constant constraint, producing a constrained saddle, and then integrates over the value of the constraint at the end. In~\eqref{E:wormholeAmplitude} the constraint is the parameter $E$, the exponential term $e^{-(\beta_1 + \beta_2)E}$ is $e^{-S}$ for the constrained saddle, $f(E)$ is the one-loop determinant at fixed constraint around the constrained saddle, and $O(G)$ encapsulates the two- and higher-loop corrections. The wormhole amplitude is then a moduli space integral, where the modulus $E$ has a critical point at the lower bound of integration $E_0$. Physically, the parameter $E$ corresponds to the asymptotic energy: these wormhole constrained saddles have two boundary energies equal to each other $E_1 = E_2 = E$, and one finds nonsingular wormholes only when $E$ is above the small black hole threshold $E_0$. In the examples we studied, $E_0$ was the black hole threshold. The wormholes are macroscopic and weakly curved when $E -E_0= O(1/G)$. However, the ``Boltzmann weight'' $e^{-(\beta_1+\beta_2)E}$ is dominated by the regime $E\to E_0$ where the wormhole pinches off; here bulk effective field theory breaks down, and the integrand in~\eqref{E:wormholeAmplitude} is inaccessible without a UV completion.  This corroborates the recent conjecture of Schlenker and Witten~\cite{Schlenker:2022dyo} that there should not exist wormholes capturing the statistics of states below the black hole threshold; indeed our wormhole contributions are only present above the black hole threshold.

Another route to~\eqref{E:wormholeAmplitude} is to study the gauge-fixed gravitational theory. Fixing a gauge, say a radial gauge in which the line element takes the form $d\rho^2 + g_{ij}dx^i dx^j$ with $x^{\mu}=(\rho,x^i)$, there are wormhole solutions to the equations of motion of the gauge-fixed theory that are not solutions to the full Einstein's equations. The wormhole is effectively supported by a constraint force coming from the gauge-fixing auxiliary field. There is a one-parameter family of these wormhole solutions labeled by the parameter $E$, and the action of these wormholes reduces to a boundary term $(\beta_1+\beta_2)E$ that depends on $E$. The integral over fluctuations around these wormholes, and over the modulus $E$, produces the amplitude~\eqref{E:wormholeAmplitude} above.

The lesson is that bulk effective field theory gives us no reliable information about the full wormhole amplitude. In some settings there is a critical point, but it is ``at infinity'' (i.e.~$E\to E_0$) where effective field theory breaks down. Presumably all manner of stringy, non-geometric contributions to the amplitude also appear at that scale. 

However, one of our main results is that if we change the question slightly, then we can stabilize the moduli and find genuine wormhole saddles. Rather than considering the two-boundary amplitude with fixed boundary metrics, consider also fixing the average energy on the two boundaries $E_{\rm avg} = \frac{E_1 + E_2}{2}$. In the bulk, the boundary energy is a component of the near-boundary metric, so this constraint may be accomplished simply through a non-standard boundary condition, where one fixes the boundary metrics and this particular combination of components near the boundary. This constraint is gauge-invariant both on the boundary and in the bulk, and so it can be consistently imposed in the quantum theory. From the point of view of the amplitude~\eqref{E:wormholeAmplitude}, the constraint has the effect of inserting a delta function $\delta(E_{\rm avg}-E)$, extracting a particular macroscopic wormhole as long as the average energy is above the minimum value $E_0$. 

As we discuss in Subsection~\ref{S:compareDoubleCone} this method for fixing the energy is related to, but not the same as the one employed by Saad, Shenker, and Stanford~\cite{Saad:2018bqo} in their study of the double cone, where they fix $E_{\rm avg}$ by an inverse Laplace transform. A crucial difference is that by directly fixing the energy we can retain boundary circles with general $\beta_1$ and $\beta_2$ and thereby study macroscopic wormholes without cone points.

Another one of our results is that $E_0$ is the small black hole threshold for a wide class of wormholes.

In boundary terms, fixing the total energy $E_{\rm tot} = E_1 + E_2$ means that we are no longer evaluating a two-replica partition function $\langle Z_1 Z_2\rangle$. Rather, we are slicing $\langle Z_1 Z_2\rangle$ by total energy, and taking a slice at fixed $E_{\rm tot}$. In an equation, we are dealing with
\beq
\label{E:constrainedZZ}
	\left\langle \text{tr}\!\left( e^{-\beta_1 H_1}\right)\text{tr}\!\left( e^{-\beta_2 H_2}\right)\right\rangle_{\mathcal{C}} \equiv \int dE_1dE_2 \,\langle \rho_1(E_1) \rho_2(E_2)\rangle \,e^{-\beta_1 E_1-\beta_2 E_2} \,\delta(E_{\rm tot} - (E_1+E_2))\,,
\eeq
where we have allowed for an average and the subscript $\mathcal{C}$ denotes that we have fixed a constraint on the total energy.

Simply put, we find a macroscopic wormhole saddle to this non-standard problem. We also accumulate evidence that a converse is true, namely that to stabilize macroscopic Euclidean wormholes (again with these boundary conditions) we must impose a constraint that relates the two boundaries.

This result already indicates something rather important. Macroscopic wormholes of this sort are dual to a non-factorizing quantity: note that $\langle Z_1 Z_2\rangle_{\mathcal{C}}$ in~\eqref{E:constrainedZZ}  does not factorize even for a single theory. As such the wormhole does not conflict with factorization.
On the other hand,  $\langle Z_1 Z_2\rangle$ for Euclidean boundaries cannot be reliably computed in bulk effective field theory; it is a UV-sensitive quantity. As a result the question of whether multi-boundary amplitudes factorize or not is, from the bulk point of view, a question about the non-perturbative definition of the bulk.

It is instructive to compare with existing results for wormhole amplitudes in JT and pure three-dimensional gravity. In both of those models wormhole amplitudes can be expressed as a moduli space integral with a critical point ``at infinity'' where the wormhole pinches off. However the UV of those models is quite different from string/M-theory, since there are no local degrees of freedom. There the moduli space integrand is completely smooth near the critical point, leading to a nonzero result for the wormhole amplitude. We expect a rather different UV behavior in stringy examples of AdS quantum gravity, like what finds for the tensionless limit of string theory with a single unit of NS-NS flux on AdS$_3\times\mathbb{S}^3\times\mathbb{T}^4$~\cite{eberhardt2021summing}.

Our macroscopic, stabilized wormhole encode physics. By analytically continuing $\beta_1 = \beta+i T$ and $\beta_2 = \beta-i T$ and taking $T\gg \beta$ we obtain a microcanonical version of the spectral form factor,
\beq
	\left\langle \text{tr}\!\left( e^{-(\beta+i T)H}\right)\text{tr}\!\left( e^{-(\beta-i T)H}\right)\right\rangle_{\mathcal{C}}\sim Te^{-\beta E_{\rm tot}},
\eeq
where $e^{-\beta E_{\rm tot}}$ is $e^{-S}$ for the analytically continued wormhole, and the factor of $T$ arises from a zero mode of the wormhole geometry where one boundary is time translated relative to the other. In writing this approximate expression we have ignored the rest of the one-loop determinant over massive fluctuations and the contribution of other topologies including the disconnected geometry.\footnote{We can modify the constraint slightly in such a way that the contribution from the disconnected geometry is also amenable to a saddle-point approximation. In addition to fixing the total energy, we can introduce a bit of Gaussian smearing in $E_1-E_2$, i.e. insert the constraint $\delta(E_{\rm tot} - (E_1+E_2)) \,e^{-\frac{(E_1-E_2)^2}{2\Delta^2}}$ into the amplitude. Then the disconnected geometry contributes $\sim e^{-\frac{\Delta^2 T^2}{2}}$ for large enough $T$, while the wormhole geometry contributes $\sim T e^{-\beta E_{\rm tot}}$.} 

This linear growth in time is a ``ramp'' in the spectral form factor, essentially the same behavior found in~\cite{Saad:2018bqo}. There are corrections to it coming from non-zero modes, higher-loop effects, and other geometries. As the wormhole gives a saddle point approximation to the form factor, we expect it to dominate over other geometries, giving a smooth ramp plus small fluctuations. This ramp only exists when the energy on each boundary is above the black hole threshold, since it is only in that domain that  the stabilized wormhole exists.

In many-body physics, the existence of a smooth ramp is a robust prediction for chaotic Hamiltonians at energies where there is a large density of states, provided that one performs some sort of averaging, whether over coupling constants as in random matrix theory, or a joint smearing in $\rho_1(E_1)\rho_2(E_2)$ over energy scales much larger than the mean level spacing.

It is not surprising then to find a ramp in gravity above the small black hole threshold, where one finds a non-perturbatively large density of unprotected states. Effectively we learn that AdS black holes have chaotic Hamiltonians. We also infer that some averaging is taking place to produce a smooth ramp, although bulk effective field theory is agnostic as to whether the disorder is description over couplings or over energies. We will comment more on this question of what averaging is taking place shortly.

In JT gravity one can study more general wormholes than the ``double trumpet,'' including geometries with more than two boundaries or where the spacetime has higher genus. One might wonder whether our methods can be used to find analogous constrained wormholes in more than two spacetime dimensions. Unfortunately even in the friendly confines of JT gravity finding these more general constrained wormholes is impossible, and so we expect failure in ordinary $d>2$ dimensional gravity as well. In JT gravity the problem is that when one considers more complicated wormholes with ``trumpets'' glued to an intermediate surface, the intermediate surface has moduli that cannot be stabilized by constraining boundary data.

\subsection{A precision test of averaging}

We put these ideas to the test to study a novel form factor. Consider a holographic CFT labeled by a coupling constant $g=g_0$ for a relevant or marginal scalar deformation $\mathcal{O}$. We have in mind a model where at $g=g_0$ the thermal one-point function of $\mathcal{O}$ is zero in two-derivative bulk effective field theory, so that the thermal entropy behaves as $S(g) = S(g_0) + O(g-g_0)^2$, and thus the mean level spacing is unaltered at the level of linear response when changing $g$. For example we may consider the Yang-Mills coupling of $\mathcal{N}=4$ super-Yang Mills, or the marginal couplings of the D1/D5 system. 

In this setting we may consider the microcanonical form factor where the two replicas have slightly different values of the coupling. These are called ``parametric correlations'' in the quantum chaos literature (see e.g.~\cite{guhr1998random} for a review). Letting $g_1 = g_0 + \delta g/2$ and $g_2 = g_0 - \delta g/2$, consider
\beq
	\widetilde{Z}=\left\langle \text{tr}\!\left( e^{- (\beta + i T)H_1}\right) \text{tr}\!\left( e^{-(\beta-i T)H_2}\right)\right\rangle_{\mathcal{C}}\,,
\eeq
where $H_i$ refers to the Hamiltonian of the dual theory at coupling $g_i$, and as above we fix the total energy $E_{\rm tot}$. There is a prediction for this quantity from the study of chaotic many-body systems analogous to the ramp~\cite{Simons:1993zza, simons1993universalities, guhr1998random}: for $\beta \ll T \ll e^S$ we have
\beq
\label{E:detunedSFF}
	\widetilde{Z}\sim T\,e^{-\pi X^2 T-\beta E_{\rm tot}}\,,
\eeq
where the rate $X^2$ is a particular one-replica average of order $\delta g^2$. This smooth exponential decay only arises upon some sort of averaging, whether over Hamiltonians or smearing over energy windows, as for the ramp.

We relate $X^2$ to a certain integrated thermal two-point function of $\mathcal{O}$ that can be computed holographically from a two-sided black hole. We also compute this form factor from a stabilized wormhole, where the bulk scalar $\phi$ dual to $\mathcal{O}$ obeys boundary conditions on the two boundaries encoding the couplings $g\pm \delta g/2$. The predicted rate $e^{-\pi X^2 T - \beta E_{\rm tot}}$ and the form factor computed by the wormhole match exactly, and in fact, from the bulk point of view, are really one and the same computation.

We regard this as a precision test of averaging in AdS/CFT, in the following sense. Usually, precision tests are quantitative matches between computations in the bulk and on the boundary. Since the boundary CFT is strongly coupled, these sorts of matches are often for supersymmetric or otherwise protected quantities. In the present instance we can only compute the thermal response function $X^2$ and the form factor from the bulk. However, the matching of these two different quantities is a prediction of a universal effective theory for chaotic systems, coming from the sigma model analysis of Altshuler and Simons~\cite{Simons:1993zza, simons1993universalities}, provided that there is some averaging. The match we perform then is a precision test, in that we are matching bulk quantities against this effective description.

\subsection{Correlations between theories with different $N$}

We also address a puzzling feature of AdS/CFT which has been recently observed by Schlenker and Witten~\cite{Schlenker:2022dyo} and has appeared in the literature at various times before (including in our previous work~\cite{Cotler:2021cqa}). One can consider wormholes connecting AdS throats supported by different amounts of RR or NS flux by putting suitable branes in between, but since the amount of flux encodes the gauge group of the dual CFT, this suggests correlations between CFTs with different gauge groups, i.e.~between different theories. This correlation may be computed using the techniques of~\cite{cotler2020gravitational, Cotler:2021cqa} and the present work for a microcanonical form factor where the two replicas have different values of $N$, in terms of a wormhole stabilized by a fixed energy constraint with a brane in the middle. The correlation is quite small, decaying exponentially in time as $\sim e^{-\gamma\frac{\partial S}{\partial N}T}$ for $S$ the entropy and $\gamma$ an order one complex number with positive real part.  Unlike the `precision test' discussed above, there is not at present a universal result in random matrix theory with which we can compare; however, it appears possible that such a result may be possible to obtain via a suitable modification of~\cite{Simons:1993zza, simons1993universalities}.

There is an analogous behavior if we consider wormholes in models where there is a gauge theory in AdS arising from compactification, but now where there is some electric flux for the Kaluza-Klein gauge fields. On the boundary we have a global symmetry, and the electric flux encodes the global symmetry charge on the boundary. The charges $Q_1$ and $Q_2$ label superselection sectors of the boundary CFT, and in the wormhole are equal by a bulk Gauss' law.\footnote{We have $Q_1=Q_2$ rather than $Q_1=-Q_2$ if the boundary orientations are endowed using outward-pointing normals from the bulk as in~\cite{Kapec:2019ecr}.} In particular the wormhole describes correlations within a superselection sector. But, by placing a charge in the middle of the wormhole, we introduce an asymmetry in the boundary charges so that $Q_1 \neq Q_2$. But this is in fact a good thing, not a puzzle, since now the wormhole allows us to compute correlations between different sectors which must be small but nonzero.

We can understand wormholes with different values of RR or NS flux on the two boundaries in a similar way, provided that we undo the decoupling limit and consider flat space string/M-theory. The Hamiltonian of flat space string/M-theory has various superselection sectors labeled by flux quanta, and the wormhole connecting throats with different RR or NS flux simply describes correlations between these different sectors.

\subsection{Comparing with the double cone}
\label{S:compareDoubleCone}

Our results with identical replicas are consistent with previous literature on the spectral form factor in AdS/CFT. Suppose that we take the extreme limit $\beta_1 = -\beta_2$, in particular $\beta_1 = i T$ and $\beta_2 = -i T$.
\footnote{While these geometries exist for general $\beta_1 = - \beta_2 = \beta+iT$, they run afoul of the recently proposed Kontsevich-Segal criterion~\cite{Kontsevich:2021dmb, Witten:2021nzp}; if $\beta = 0$ we can satisfy the Kontsevich-Segal criterion, although we need to smooth out the cone point with a complex radial diffeomorphism. It should be noted that matter propagation in the 2d version of the geometry with $\beta_1 = - \beta_2$ is well-defined for $\beta \neq 0$ as long as $T\neq 0$.}
In this limit there are genuine wormhole saddles of Einstein gravity, the ``double cones'' of~\cite{Saad:2018bqo}. These geometries are orbifolds of the exterior of a two-sided black hole by a translation in Killing time, joining two Lorentzian cones at the tips. There is a family of these wormholes characterized by the same quantities that label a black hole -- namely its energy, angular momentum, and charge -- which are the same in both exteriors. The action of the double cone vanishes identically for all values of these parameters, i.e.~these quantities are flat directions at tree level. Following a similar analysis for the SYK model and in JT gravity performed in~\cite{Saad:2018bqo}, we can select a particular wormhole of total energy $E_{\rm tot}$ by taking $\beta_1 = \beta + i T$ and $\beta_2 = \beta - i T$, and inverse Laplace transforming in $\beta$ with a Gaussian window, or in an equation,
\beq
	\includegraphics[width=5.3in, valign = c]{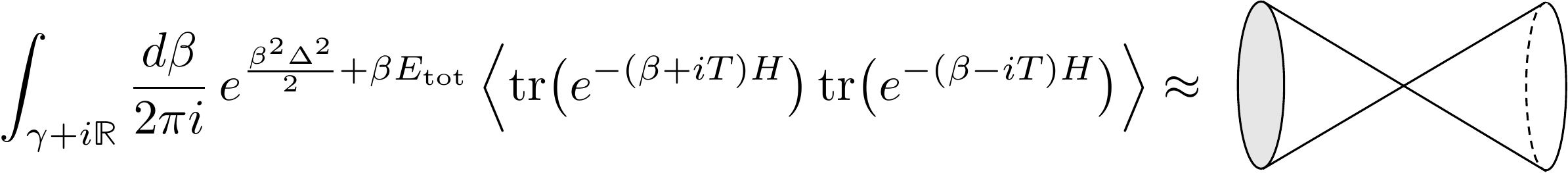}
\eeq
This quantity has a saddle point approximation at $\beta = 0$, the double cone itself, at boundary energy $E_1 = E_2 =E_{\rm tot}/2$.\footnote{The Gaussian smearing ensures that the integral over $\beta$ has a saddle-point approximation.} We could also Fourier transform in (angular) chemical potentials to fix the angular momentum and charge. So to select a particular double cone we must fix the total energy. As in our discussion above, even if the bulk is dual to single instance, the common energy constraint ensures that the dual of the double cone of a fixed energy is not a factorizing quantity, and it does not necessarily imply a breakdown of factorization.

The double cone then, after inverse Laplace transforming to fixed energy, also gives a microcanonical form factor, essentially the same one computed by directly fixing the energy, except now at $\beta = 0$. We do note though that, when we fix the energy directly rather than by an integral transform, the dual geometry does not have timelike cone points to regulate.

We observe that one can pick out the double cone via a constraint that does not couple the two replicas.  In particular, one can obtain a particular double cone by fixing the energy on a single boundary without having to pick a non-local constraint that involves both boundaries. Then the double cone would lead to a genuine factorization paradox, in the sense that there is a factorizing quantity on the boundary which is dual to a non-factorizing quantity in the bulk.  It is possible that embedding the double cone into string theory could lead to instabilities, the most likely candidate being light strings or branes wrapping the shrinking time circle. If the double cone is a sensible configuration in AdS string theory then perhaps UV physics cures the factorization paradox in this setting.  In light of these subtleties, we will mostly study Euclidean wormholes, and more generally complex wormholes where the bottleneck does not pinch off, in which some of these matters are under better control.

The story for the spectral form factor is richer when the two replicas have slightly different values of the coupling $g$. Now there is a background for a scalar field $\phi$ in the bulk dual to $\mathcal{O}$. Fixing the energy directly, we can retain $\beta>0$ and our wormholes remain macroscopic. However, if we instead perturb around the double cone, we run a problem. The scalar background moves the saddle point value of $\beta$ to negative values, $\beta \sim - \delta g^2 T$. The ensuing perturbed double cone violates the recently proposed Kontsevich-Segal criterion~\cite{Kontsevich:2021dmb, Witten:2021nzp} for admissible complex metrics, i.e.~the action for matter fields in this background does not have a real part which is bounded below.

There may yet be some way to understand the form factor with detuned couplings using the double cone. But, for now, this is an instance where the machinery we have recently developed in~\cite{cotler2020gravitational, Cotler:2021cqa}, combined with a direct constraint on boundary energy, allows us to compute the form factor~\eqref{E:detunedSFF} without violating the Kontsevich-Segal criterion. And, we note again that this result works for a wide range of bulk effective field theories. At once we gain confidence in our methods and learn about averaging in AdS/CFT.

\subsection{Where do we go from here?}

To summarize:
\begin{enumerate}
	\item We find wormhole saddles to a non-standard problem in AdS quantum gravity connecting two $\mathbb{S}^1\times\mathbb{S}^{d-1}$ or torus boundaries in bulk effective field theory by constraining the total boundary energy. The wormhole does not violate factorization on account of the constraint.
	\item These saddles compute a microcanonical version of the spectral form factor, giving a smooth ramp with small fluctuations as in~\cite{Saad:2018bqo}.
	\item We show that a wide family of wormholes have energies only above the  small black hole threshold. 
	\item We used these methods to compute a form factor where the two replicas have slightly different coupling constants. The form factor decays at a rate which exactly matches a one-replica prediction from the many-body literature. The decay is smooth with small fluctuations, like the ramp when the boundary couplings are equal. 
	\item  The smoothness of the ramp and, when the couplings are detuned, of the decay, indicates some averaging.
	This averaging is visible in bulk effective field theory only above the black hole threshold, and bulk effective field theory is agnostic as to whether the averaging is over Hamiltonians or a smearing over energies.
	\item We find wormhole saddles with branes which describe small correlations between holographic CFTs with different values of $N$.
\end{enumerate}

A punchline of our work is that, at least as far as these wormholes are concerned, a bulk effective field theorist can detect averaging in the black hole spectrum, but cannot distinguish between averaging over coupling constants or smearing in energies. Relatedly, our bulk computations
cannot distinguish whether AdS black hole Hamiltonians are random or pseudo-random.

This result is reminiscent of another, older subject in theoretical physics, namely semiclassical quantum chaos.  In this literature, effective descriptions of smeared level statistics as well as smeared bands of eigenstates can be computed for quantum chaotic systems using classical particle trajectories (e.g.~\cite{berry1972semiclassical, heller1984bound, berry1985semiclassical, casati2006quantum, gutzwiller2013chaos, heller2018semiclassical}).  As the energy smearing window is taken to zero size, the semiclassical description becomes more complicated since progressively more intricate trajectories need to be included (for instance, periodic orbits with very large period) to maintain the reliability of the approximation.  In the extreme limit that we hone in on individual energy levels or individual eigenstates, the semiclassical approximation ceases to exist entirely.  Similar techniques are used in the study of mesoscopic systems (see~\cite{simons2002mesoscopic} for a review), which often are many-body.  A mescoscopic description of a system is one which expresses statistical properties of the underlying UV physics, often via energy smearing or disorder as a proxy.

With this history in mind, we would like to draw an analogy between the mesoscopic description of chaotic systems and bulk effective field theory. Namely, we regard semiclassical gravity as a mesoscopic theory of spacetime, able to describe coarse-grained aspects of black hole physics but not the precise details of the microstate spectrum, spectral correlations, etc. (A related outlook has been suggested in~\cite{Pollack:2020gfa}, in the context of the eigenstate thermalization hypothesis.) This might be the best we can do in practice without the computational resources (classical or quantum) required to directly access individual energies and eigenstates in a boundary dual; moreover these `exact' computations are bound to involve a high degree of complexity both in their implementation and in the description of their outputs.  
And, even if we had the exact spectral data of a holographic CFT, relating this to a bulk computation would require a non-perturbative holographic dictionary and an independent non-perturbative definition of the bulk, both of which are presently lacking.

The above being said, mesoscopic quantum gravity may lead to many new insights, as well as novel tools for analyzing the quantum structure of spacetime in AdS as well as other cosmologies.  For instance, there is promise to port tools from semiclassical quantum chaos to obtain descriptions of energy-smeared bands of black hole microstates (for recent work in this direction, see~\cite{Dodelson:2022eiz}); indeed we have initiated this program for the form factor in the present paper.  In retrospect, many of the developments in quantum gravity over the last several years surrounding replica and Euclidean wormholes wormholes can be thought of as arising from a mesoscopic description of spacetime. We emphasize, however, that new techniques (such as gravitational constrained saddles~\cite{cotler2020gravitational, Cotler:2021cqa} and $\lambda$-solutions~\cite{Cotler:2021cqa}) are often required to fully realize analogues of these results in $d > 3$ dimensions, and that the generalizations need not be direct analogues of the lower-dimensional results. In particular, there need not be intrinsic disorder in higher dimensional gravity, but rather a smearing over energies.

We have gone out of our way to note that while our results for wormholes and form factors imply some averaging, the nature of the averaging is an unanswerable question in bulk effective field theory. However, there is strong reason to believe that in stringy AdS/CFT the averaging is over energy windows and not Hamiltonians, and relatedly, that the black hole Hamiltonian is pseudo-random rather than random.

Let us argue for this statement via the contrapositive. Suppose that holographic CFTs dual to stacks of D- or M-branes are disordered. In the vacuum, bulk effective field theory preserves boundary conformal invariance to all orders in the bulk coupling constants, implying that any distribution for dimensionful coupling constants must be non-perturbatively small at large $N$. As for marginal couplings, these do not exist in $d>4$ superconformal field theories with a gravity dual, and even an average over say the complexified Yang-Mills coupling of $\mathcal{N}=4$ super-Yang Mills like that studied in~\cite{Collier:2022emf}, while of independent interest, seems to be rather different than what one finds in string theory on AdS$_5\times\mathbb{S}^5$. The problem is the large set of precision tests involving operators and defects dual to strings and branes that depend sensitively on the Yang-Mills coupling, like the dimension $\Delta \sim \sqrt{\lambda}$ of the Konishi operator~\cite{Gromov:2009zb}, circular 1/2-BPS Wilson loops on $\mathbb{S}^4$ (e.g.~\cite{Erickson:2000af}) which in the fundamental representation go as $\langle W\rangle \sim \frac{e^{\sqrt{\lambda}}}{\lambda^{3/4}}$, central charges associated with 1/2-BPS surface defects which behave as $\frac{N}{\lambda}$~\cite{Jensen:2018rxu,Chalabi:2020iie}, and many more. All of these depend on the Yang-Mills coupling in rather different ways. If the holographic dictionary involves an average over the Yang-Mills coupling, with the boundary value of the dilaton encoding the average value of the Yang-Mills coupling, then the distribution in couplings must be very narrow indeed to not run afoul of these tests. The distribution would have to satisfy e.g. $\langle \sqrt{\lambda}\rangle \approx \sqrt{\langle\lambda\rangle}$, $\left\langle \frac{e^{\sqrt{\lambda}}}{\lambda^{3/4}}\right\rangle \approx \frac{e^{\sqrt{\langle \lambda\rangle}}}{\langle \lambda\rangle^{3/4}}$, $\left\langle \frac{N}{\lambda}\right\rangle \approx \frac{N}{\langle \lambda\rangle}$, etc., to the precision of existing holographic results.

There are also clues from the bulk. As we have stressed, the smooth macroscopic wormholes that encode spectral correlations only exist above the small black hole threshold. It is precisely at this threshold where the density of states becomes non-perturbatively large in the Newton's constant, and one can have self-averaging of coarse-grained quantities like a smeared version of the spectral form factor. Furthermore, there are some indications that wormholes do not contribute to protected quantities. For example, it has been explicitly demonstrated in supersymmetric versions of JT gravity that multi-replica partition functions factorize when taking boundary conditions corresponding to the superconformal index on each boundary~\cite{Iliesiu:2021are}. There the connected geometries have gravitino zero modes that sets those contributions to vanish, and one can envision a similar infrared mechanism for more general protected quantities.\footnote{For example one could consider wormholes that contribute to the two-replica version of the index for $\mathcal{N}=4$, stabilized by a fixed energy constraint as in our analysis.} It is a tall order to have an ensemble averaging in the boundary which, to good approximation, is only over the spectrum of unprotected black hole microstates while leaving protected quantities alone. Such an average would approximately preserve the full superconformal symmetry, which seems to be impossible in theories without a conformal manifold like the $\mathcal{N}=(2,0)$ theory, and as we mentioned above would be constrained by precision tests involving strings and branes in models with marginal couplings. 

In contrast, there is no tension between pseudo-randomness in the very dense spectrum of unprotected black hole microstates and a lack of averaging in the protected part of the spectrum. Indeed, pseudo-randomness is the norm rather than the exception in many-body quantum systems, and from that point of view we should expect pseudo-randomness in the unprotected spectrum of holographic CFTs. Furthermore, bulk effective field theory produces results in a power series in $1/N$, and so perturbation theory around known saddles is incapable of resolving physics on the scale of the mean level spacing of black hole microstates, of $O\!\left( e^{-1/G}\right)$. It seems quite reasonable then that the bulk effective theory is smearing over those scales.  In particular, a smearing over a window of size $N^m \exp\!\left( -\frac{1}{G}\right)$ for $m \sim O(1)$ would be much larger than the mean level spacing, but invisible in bulk effective field theory, being non-perturbatively small at large $N$. This logic is similar to that appearing in~\cite{Schlenker:2022dyo}, although here we emphasize energy smearing as an essential byproduct of the inability to resolve scales of order the level spacing.

In advocating for pseudorandomness in AdS/CFT rather than genuine randomness, there is a fly in the ointment that we have deferred discussing until now. Namely, what do we do about genuine wormhole saddles of supergravity in AdS? If such stable wormhole saddles exist, without introducing constraints, then they would imply a breaking of factorization accessible in bulk effective field theory. This is not a problem a priori, since contributions from a stringy UV completion could restore factorization.  However, it would be ideal if factorization holds even in effective field theory.

The taxonomy of Euclidean AdS wormhole saddles is nicely reviewed in the Introduction of~\cite{marolf2021ads}, and fortunately it is rather short. In particular, most known saddles have been argued to be unstable one way or another, whether under the nucleation of brane pairs as in uplifts of the higher genus wormholes in~\cite{maldacena2004wormholes},\footnote{It is worth noting that the AdS$_3$ wormholes of~\cite{Belin:2020hea, Chandra:2022bqq} include these higher genus wormholes, and~\cite{Chandra:2022bqq} also contains a study of wormholes with lower genus boundaries supported by heavier operators whose conformal dimensions scale with the central charge $c$, dual to conical deficits. The former are known to be unstable to nucleation when embedded into string theory, while the status of the latter is uncertain for a few reasons. (i) One requires AdS$_3$ vacua of string theory with heavy extended objects in AdS$_3$, much heavier than that of strings and D-branes. (ii) If one instead tries to support a wormhole with a large number of insertions of strings or branes, the finite cross-section and the Gauss' laws for NS and RR flux would imply a distribution with an equal number of branes and anti-branes suspended between the boundaries, and it is unclear if such a stable configuration exists. (iii) If one does embed such a wormhole into string theory, it may yet be unstable against nucleation.
That being said the detailed match between wormhole saddles of \textit{pure} three-dimensional gravity and a simple ensemble of CFT data seems too remarkable to be a coincidence, and despite lying outside the standard paradigm of AdS/CFT it deserves further study.} or with respect to perturbative fluctuations.
To our knowledge there are only three wormholes in AdS supergravity that have not yet been argued to be unstable: the asymptotically $\mathbb{H}_5\times\mathbb{S}^5$ wormhole of Maldacena and Maoz connecting two $\mathbb{S}^4$ boundaries, the asymptotically $\mathbb{H}_3\times\mathbb{S}^3\times \mathbb{T}^4$ axion wormhole of Arkani-Hamed, Orgera, and Polchinski~\cite{Arkani-Hamed:2007cpn} connecting two $\mathbb{S}^2$ boundaries, and an asymptotically $\mathbb{H}_4\times\mathbb{S}^7$ wormhole of Marolf and Santos connecting two $\mathbb{S}^3$ boundaries. A complete analysis of potential perturbative instabilities has been performed for the Marolf-Santos wormhole (although the details did not appear in their paper), and only a partial analysis has been performed for the Maldacena-Maoz wormhole. The simplest channel for brane nucleation instabilities was also checked for the Marolf-Santos wormhole.

A notable Lorentzian wormhole saddle is, of course, the double cone of Saad, Shenker and Stanford~\cite{Saad:2018bqo}.  As we remarked above, this solution does not necessarily require a joint constraint between the two boundaries, and so appears to be a genuine semiclassical saddle contribution to a factorizing CFT quantity.  However, as mentioned above, there are some extant questions about the double cone concerning its status in string theory. 
In particular strings and branes wrapping the time circle can become light near the tip of the cone, potentially precluding a semiclassical description.

It seems to us that these wormholes demand further study. It is important to conclusively address whether they are stable saddles or not in string theory. If they are, perhaps there is a subtle way in which the wormhole involves a non-local constraint between the two boundaries, in such a way that the wormhole does not contradict factorization.  If some are not, then we may require UV physics to restore factorization.

The remainder of this manuscript is organized as follows. In the next Section we review a few details about wormhole amplitudes in AdS/CFT. Then, in Section~\ref{sec:fixing} we introduce a fixed energy constraint, and show that a wide class of wormholes exist only above the black hole threshold.  In Section~\ref{sec:heavier} we show that the wormholes accounting for black hole microstate energy correlations only exist at energies above the black hole threshold.  We go on to study the form factor between replicas with slightly different coupling constants in Section~\ref{sec:gcorr}, and different values of $N$ in Section~\ref{sec:Ncorr}.

\section{Some review about wormholes in AdS}
\label{sec:review}

In this section we review our previous work~\cite{cotler2020gravitational,Cotler:2021cqa} about wormholes in AdS$_{d+1}$ where the boundary is $\mathbb{S}^1\times \mathbb{S}^{d-1}$ or $\mathbb{T}^d$.  We discuss wormholes in Einstein gravity, but the methods here may also be applied to AdS compactifications of 10- and 11-dimensional supergravity~\cite{Cotler:2021cqa}.

It has been shown by Witten and Yau~\cite{witten1999connectedness} that there are no Euclidean wormhole solutions with two asymptotically Euclidean AdS boundaries with non-negative curvature, like $\mathbb{S}^1\times\mathbb{S}^{d-1}$ or $\mathbb{T}^d$.  So without turning on sources for bulk matter fields, the wormholes that contribute to multi-boundary amplitudes with those asymptotics are always off-shell.

Nevertheless we would like to see if we can say something reliable about wormhole amplitudes in AdS. In quantum field theory there is machinery designed to deal with this sort of problem, namely the method of constrained instantons~\cite{affleck1981constrained}, which we will refer to as constrained saddles. This has famously been used to compute instanton-like corrections in the Higgs phase of gauge theories where no instanton saddles exist~\cite{affleck1984dynamical}.  The basic idea of constrained saddles is to introduce a delta-functional constraint $\delta(\mathcal{C}[\varphi] - \zeta)$ in a path integral over fields $\varphi$, find a saddle at each fixed value $\zeta$ of the constraint, and then integrate over all such saddles at the end.  In our previous work~\cite{cotler2020gravitational}, we adapted this method to find two-sided wormhole contributions to pure Einstein gravity, where we constrained the length between the two boundaries.  However, there is a better method from our later paper~\cite{Cotler:2021cqa} which we will rely on for the remainder of the manuscript.

The basic idea is as follows. In the gravitational path integral we divide by diffeomorphisms, and even classical gravity is only well-defined upon gauge-fixing. We can then look for wormhole solutions to the equations of motion of the gauge-fixed theory, which form a subset of the full Einstein's equations. For the purpose of finding Euclidean wormholes with two asymptotic regions, it is convenient to work in coordinates $(\rho, x^i)$ where $\rho$ will ultimately be the `radial' direction along the wormhole. It is useful to fix a radial gauge
\begin{equation}
	g_{\mu \rho} = \delta_\mu^\rho\,,
\end{equation}
which can be implemented by adding a gauge-fixing term to the gravitational action,
\begin{equation}
S_{\text{gauge-fixing}} = i \int d^{d+1}x \, \sqrt{g} \, \lambda^\mu(g_{\mu \rho} - \delta_\mu^\rho)\,,
\end{equation}
where $\lambda^\mu$ is the gauge-fixing auxiliary field. The gauge-fixed line element is
\beq
	ds^2 = d\rho^2 + g_{ij}(\rho,x^k)\,dx^idx^j\,,
\eeq
and the equations of motion of the gauge-fixed theory are simply the $ij$ components of the Einstein's equations,
\begin{equation}
\label{E:lambdaEOM1}
	R_{ij}-\frac{R}{2}\,g_{ij}+\Lambda g_{ij} = 0\,.
\end{equation}
In effect, the radial components of Einstein's equations may be violated in the gauge-fixed theory, soaked up by a profile for the auxiliary field $\lambda^{\mu}$, which acts as a constraint force. Of course usually what we do when studying Einstein gravity is look for solutions to the full Einstein's equations, with no constraint force.  
But as far as the path integral is concerned, we can also find saddles with a non-zero profile for $\lambda^\mu$, and so long as we sum over all such solutions the final result for an amplitude ought to be gauge-invariant. 

In~\cite{Cotler:2021cqa} we termed solutions of this sort ``$\lambda$-solutions'', and
remarkably, there are non-trivial $\lambda$-solutions to~\eqref{E:lambdaEOM1} describing Euclidean AdS wormholes connecting two asymptotic regions~\cite{Cotler:2021cqa}.  For instance, for asymptotically unit radius $\mathbb{H}_{d+1}$ spaces with boundaries $\mathbb{S}_{\beta_1}^1 \times \mathbb{T}^{d-1}$ and $\mathbb{S}_{\beta_2}^1 \times \mathbb{T}^{d-1}$, we have the family of solutions
\beq
\label{E:torussols1}
ds^2 = d\rho^2 + \frac{b^2}{4}\left(2 \, \cosh\left(\frac{d\,\rho}{2}\right)\right)^{\frac{4}{d}} \left(\left(\frac{\beta_1 \, e^{\frac{d\,\rho}{2}} + \beta_2 \, e^{- \frac{d \, \rho}{2}}}{2 \, \cosh\left(\frac{d\,\rho}{2}\right)}\right)^2 d\tau^2 + d\vec{x}_\perp^2\right)\,,
\eeq
parameterized by a modulus $b > 0$.  Here $\tau\sim\tau+1$ is the coordinate along the Euclidean time circle and $\vec{x}_{\perp}$ describes a spatial torus.
In five dimensions with unit radius and boundaries $\mathbb{S}_{\beta_1}^1 \times \mathbb{S}^3$ and $\mathbb{S}_{\beta_2}^1 \times \mathbb{S}^3$ we similarly have the family of solutions
\beq
\label{E:S1S3sol1}
ds^2 = d\rho^2 + \frac{b^2 \cosh(2\rho) - 1}{2} \left(\left(\frac{\beta_1 \, e^{2\rho} + \beta_2 \, e^{-2\rho}}{2\left(\cosh(2\rho) - \frac{1}{b^2}\right)}\right)^2 d\tau^2 + d\Omega_3^2\right)\,, 
\eeq
parameterized by a modulus $b>1$. We emphasize that~\eqref{E:torussols1} and~\eqref{E:S1S3sol1} are not solutions to Einstein's equations, but rather are solutions to~\eqref{E:lambdaEOM1} of the gauge-fixed theory.

Our $\lambda$-solutions in~\eqref{E:torussols1} and~\eqref{E:S1S3sol1} are each labeled by a parameter $b$, which controls the size of the bottleneck of the wormholes.  In each case, $b$ can also be interpreted as boundary energy, and the action of the wormholes depend on this parameter~\cite{cotler2020gravitational, Cotler:2021cqa}.\footnote{How can this happen? Normally the action of a solution is stationary against perturbations. What happens here is that fluctuations in $b$ change the boundary metric, but by a constant conformal rescaling which is consistent with the AdS boundary conditions.  This leads to a boundary term in the variation.}  For instance, for the $\mathbb{S}^1 \times \mathbb{T}^{d-1}$ wormholes for which $b > 0$, the boundary energy is
\begin{equation}
E = \frac{\text{Vol}(\mathbb{T}^{d-1})\, (d-1)\, \left(\frac{b}{2}\right)^d}{4 \pi G}\,,
\end{equation}
and the gravitational action at fixed $b$ (and thus fixed $E$) is simply
\begin{equation}
S_{\text{grav}}^{\mathbb{S}^1 \times \mathbb{T}^{d-1}} = (\beta_1 + \beta_2) E\,.
\end{equation}
Here the energy is in the range $[0,\infty)$.  Similarly, for the $\mathbb{S}^1 \times \mathbb{S}^3$ wormholes for which $b > 1$, the boundary energy is
\begin{equation}
E = b^4 E_0 = \frac{3\pi b^4}{32 G}
\end{equation}
and likewise the gravitational action at fixed $b$ (or $E$) is
\begin{equation}
S_{\text{grav}}^{\mathbb{S}^1 \times \mathbb{S}^{3}} = (\beta_1 + \beta_2) E\,.
\end{equation}
In this case, the energy is in the range $[E_0, \infty)$, where $E_0$ is the energy of the lightest non-rotating black hole with $\mathbb{S}^1 \times \mathbb{S}^3$ boundary topology in Euclidean signature.

In light of the above, for fixed topology we should integrate over the modulus $b$ (really the boundary energy) in the path integral.  This means integrating over our families of $\lambda$-solutions.  This produces an integral representation of the wormhole amplitude which can be written as~\cite{cotler2020gravitational, Cotler:2021cqa}
\begin{equation}
\label{E:wormholeschematic2}
Z_{\text{wormhole}}(\beta_1, \beta_2)  = \int_{E_0}^\infty dE \, e^{-(\beta_1 + \beta_2) E} f(E; \beta_1, \beta_2) (1 + O(G))\,,
\end{equation}
where $E_0$ is the energy of the lightest non-rotating black hole with the same boundary topology as the wormhole.  It is natural to ask: what would happen if we had picked another gauge?  We would have gotten a different set of gauge-fixed equations of motion and correspondingly different $\lambda$-solution geometries, but the integrated amplitude $Z_{\text{wormhole}}(\beta_1, \beta_2)$ would still be the same.

When it comes to evaluating~\eqref{E:wormholeschematic2}, we run into a problem.  The integral is dominated by regime where the bottleneck shrinks to zero, corresponding to $E\to E_0$.  This means that our line of $\lambda$-solutions terminates in a singular wormhole. For torus boundaries, this wormhole is a critical point ``at infinity.'' On account of the bottleneck shrinking to zero size, in $d+1 > 3$ where there are propagating degrees of freedom in the bulk, the amplitude is UV sensitive and so bulk EFT cannot tell us about the full integrated amplitude in~\eqref{E:wormholeschematic2}.

Fortunately there is a way out of this impasse, as we discussed in previous work~\cite{Cotler:2021cqa}.  For our $\lambda$-solutions in~\eqref{E:torussols1} and~\eqref{E:S1S3sol1}, we can consider the continuation $\beta_1 = \beta + i T$, $\beta_2 = \beta-iT$ in order to study the spectral form factor at finite inverse temperature $\beta$ and time $T$.  Then our wormholes become genuine (albeit complex-valued) saddles of the Einstein equations for $\beta = 0$.  In fact, the geometry of each saddle is precisely the double cone of Saad, Shenker and Stanford~\cite{Saad:2018bqo} with the appropriate boundary topology.  In each case, the saddle has tree-level moduli, including the energy $E$ and other charges.  To fix them one can inverse Laplace transform in the energy $E$, which amounts to an integral over $\beta$.  This integral over $\beta$ also has saddle-point approximation provided we introduce a suitable Gaussian broadening~\cite{Saad:2018bqo, Cotler:2021cqa}.

The resulting spectral form factor amplitude is unconventional -- we had to take an inverse Laplace transform after all -- but we do end up with a non-standard wormhole saddle of Einstein gravity.  This program also works in the supergravity compactifications relevant for the basic examples of AdS/CFT~\cite{Cotler:2021cqa}, and moreover interplays in a surprisingly nice manner with brane effects~\cite{Cotler:2021cqa, Mahajan2021wormholes}.

In the next section we discuss a different way, other than the inverse Laplace transform procedure, to get wormholes as saddles to non-standard questions in AdS gravity.  Our approach here will clarify the meaning and role of the wormholes in the holographic dictionary.

\section{Fixing the energy and averaging}
\label{sec:fixing}

In the Introduction and the previous Section we reviewed some facts about wormholes in JT and AdS$_3$ gravity, as well as in higher-dimensional Einstein gravity. Crucially, there is a breakdown of factorization across boundaries in JT gravity, and strong evidence for such a breakdown in AdS$_3$ gravity as well. In higher-dimensional gravity wormholes contribute to a ramp in the spectral form factor through the double cone, and this seemingly violates factorization as well, insofar as wormholes contribute to the connected part of the two-boundary amplitude, which is the very thing that must vanish if multi-boundary amplitudes factorize.  

It would be useful to have a non-singular version of the double cone, and more generally stabilized versions of purely Euclidean wormholes. In this Section we show how to obtain such non-singular wormholes by introducing a further boundary condition in the sum over metrics, rather than by performing an inverse Laplace transform. From the Euclidean wormhole amplitude in Eq.~\eqref{E:wormholeschematic2},
\begin{equation*}
	Z_{\rm wormhole}(\beta_1,\beta_2) = \int_{E_0}^{\infty} dE\, e^{-(\beta_1+\beta_2)E}f(E; \beta_1, \beta_2)\left( 1 + O(G)\right)\,,
\end{equation*}
we might hope that all we need to do to find wormhole saddles of Euclidean gravity is to somehow fix the quantity $E$ directly, in addition to $\beta_1$ and $\beta_2$. In this Section we find a boundary condition in AdS gravity that directly fixes boundary energy, and show that it can be used to stabilize the Euclidean wormholes we presented in the last Section. 

\subsection{Fixing the energy in field theory}

Consider a quantum field theory in flat space at finite temperature. Expressing the partition function as a sum over states,
\beq
	Z(\beta) = \int dE\, \rho(E) \,e^{-\beta E}\,,
\eeq
if we insert a delta function $\delta (E-E')$ into this sum we extract the integrand at energy $E'$, $\rho(E')\,e^{-\beta E'}$, essentially the density of states. There is a simple way to introduce this constraint in field theories with a two-derivative Lagrangian description.

Ordinarily we would extract the density of states $\rho(E)$ by inverse Laplace transform,
\beq
	\rho(E) = \int_{\gamma+i \mathbb{R}} \frac{d\beta}{2\pi i} \,e^{\beta E}Z(\beta)\,,
\eeq
but for our purposes it will be useful to be able to directly study physics at energy $E$ without introducing an intermediate integral over $\beta$.

The thermal partition function can be written as a functional integral over fields on a Euclidean space $\mathbb{S}^1_{\beta} \times \mathcal{M}$. This path integral may be written in Hamiltonian form by integrating in momenta $p_a$ conjugate to the original fields $q^a$, and then performing
\beq
	Z = \int [dq][dp] \exp\left(  \int_0^{\beta} d\tau \left( ip_a \partial_{\tau} q^a- H(q,p)\right)\right)\,,
\eeq
perhaps subject to constraints. Now we simply insert a delta function
\beq
	\delta\! \left(E'- \frac{1}{\beta}\int_0^{\beta} d\tau \,H(q,p) \right)\,,
\eeq
into the functional integral. Formally the constrained integral reads
\beq
	\widetilde{Z}(\beta;E) = e^{-\beta E'}\int [dq][dp] \exp\left( i \int_0^{\beta} d\tau \,p_a \partial_{\tau} q^a\right) \delta\!\left( E'-\frac{1}{\beta} \int_0^{\beta} d\tau \,H\right) 
\eeq
Now observe that we can scale $\beta$ out of the residual integral, so that the constrained quantity reads
\beq
	e^{-\beta E'} \int [dq] [dp] \exp\left( i \int_0^1 d\tau \, p_a\partial_{\tau} q^a \right) \delta\!\left( E'-\int_0^1d\tau H(q,p)\right)\,.
\eeq
But this is nothing more than the desired $\rho(E')\,e^{-\beta E'}$. In fact this leads to a formal functional integral for the density of states, 
\beq
	\rho(E) = \int [dq][dp] \exp\left( i \int_0^1 d\tau\, p_a\partial_{\tau} q^a\right) \delta\!\left( E-\int_0^1 d\tau\,H(q,p)\right)\,,
\eeq
although in practice this path integral is impossible to perform as it lacks a saddle-point approximation.

The virtue of passing over to Hamiltonian form is that it is manifest that the periodicity $\beta$ drops out of this path integral for $\rho(E)$. Of course what we have done is equivalent to the inverse Laplace transform, and it therefore holds for thermal field theory more generally. Remaining within the Hamiltonian framework, we can write the delta function constraint using a Lagrange multiplier $\lambda$, 
\beq
	\delta\!\left( E-\int _0^{1}d\tau\,H(q,p)\right) = \int_{i\mathbb{R}} \frac{d\lambda}{2\pi i} \,\exp\!\left( \lambda\left( E-\int_0^1 d\tau \,H\right)\right)\,,
\eeq
so that upon relabeling $\lambda =i\tilde{\beta}$, rescaling time again as $\tau \to \frac{\tau}{\tilde{\beta}}$, and pushing the integration contour so that the $\lambda$ integral converges, we have
\begin{align}
\begin{split}
	\rho(E)& = \int_{\gamma+i \mathbb{R}} \frac{d\tilde{\beta}}{2\pi i}\,e^{\tilde{\beta}E}\int [dq][dp] \exp\!\left( i \int_0^{\tilde{\beta}}d\tau \,p_a \partial_{\tau} q^a- \int_0^{\tilde{\beta}}d\tau\,H(q,p)\right)
	\\
	& = \int_{\gamma+i\mathbb{R}}\frac{d\tilde{\beta}}{2\pi i}\,e^{\tilde{\beta}E} Z(\tilde{\beta})\,,
\end{split}
\end{align}
recovering the inverse Laplace transform as expected.

In holographic CFTs the partition function $Z(\beta)$ has a saddle-point approximation, as does the integral over $\beta$ that computes the density of states. When that is the case, the direct energy constraint is more of a hindrance than a help. By imposing a constraint on an integral that already has a saddle-point approximation, we produce an over-constrained system. The constraint will prove its worth when we consider wormhole amplitudes, which lack a saddle-point approximation and for which we will want to hold the boundary data $\beta$ fixed.

\subsection{Fixing the energy in gravity}

Now we would like to fix the energy in AdS gravity, following in the footsteps of~\cite{Brown:1992bq, Brown:1993ke, Marolf:2018ldl, Marolf:2022jra}. Let us first consider the one-boundary problem. One of the more basic entries in the holographic dictionary is the relation between the boundary stress tensor and the bulk metric~\cite{balasubramanian1999stress},
\beq
	T^{\mu\nu} = \lim_{\rho\to\infty} \left[ \frac{\sqrt{\gamma}}{8\pi G \sqrt{g^{(0)}}}\left( K^{\mu\nu} - K \gamma^{\mu\nu}\right)+T^{\mu\nu}_{\rm CT}\right]\,,
\eeq
where $\rho$ is a radial coordinate that tends to infinity at the conformal boundary, $\gamma$ is the metric on a cutoff slice of constant $\rho$, $g^{(0)}_{\mu\nu}$ is the boundary metric to be defined shortly, $K_{\mu\nu}$ is the extrinsic curvature of that cutoff slice, and $T^{\mu\nu}_{\rm CT}$ is a counterterm that arises from holographic renormalization. For example, in pure gravity, we may pick coordinates near a boundary reached as $\rho\to\infty$ so that the line element reads\footnote{The reason we have the clause about pure gravity is that the precise details of the near-boundary expansion depend on whatever sources we turn on for operators dual to bulk matter fields.}
\beq
	ds^2 = d\rho^2 + e^{2\rho} \left( g^{(0)}_{\mu\nu}(x) + e^{-2\rho} g^{(2)}_{\mu\nu}(x) + \hdots +e^{-d\,\rho} \,\rho h_{\mu\nu} (x)+ e^{-d\,\rho} g^{(d)}_{\mu\nu}(x) + \hdots\right)dx^{\mu}dx^{\nu}\,,
\eeq
where $g^{(0)}_{\mu\nu}$ is the boundary metric which is fixed as a boundary condition, the subleading terms $g^{(2)}$ up through $h$ are fixed by the boundary metric through the Einstein's equations, and $g^{(d)}$ as well as further subleading terms are allowed to fluctuate in the sum over metrics. The fluctuating part of the boundary stress tensor is completely determined by the subleading component $g^{(d)}_{\mu\nu}$. 

Boundary energy density is a gauge-invariant quantity, and we can introduce a non-standard boundary condition in AdS gravity by fixing not only the boundary metric, but also the integrated energy density, where we have in mind Euclidean boundaries of the form $\mathbb{S}^1\times\mathcal{M}$. 

What precisely is being fixed in the dual description? Since the bulk metric fluctuates, it is reasonable to think that this boundary condition fixes the integrated energy density $\frac{1}{\beta}\int d^dx \sqrt{g^{(0)}} \, T^{\tau}{}_{\tau}$ in the boundary theory; that is, this boundary condition fixes microscopic energy in the dual.

If this is indeed the case, the sum over metrics with this boundary condition will produce $\rho(E')\,e^{-\beta E'}$, just as in our discussion in the last Subsection. Rewriting Einstein gravity in the ADM formalism, i.e.~using a Hamiltonian framework in the bulk, the dynamical fields are the spatial metric $g_{ij}$ and its conjugate momentum $\pi^{ij}$, and the sum over metrics can be written as
\beq
	\int [dg][d\pi] \,\exp\!\left( i\int d^{d+1}x \,\pi^{ij}\partial_{\tau} g_{ij} -\int d^dx \sqrt{g^{(0)}}\,\mathcal{H}(g,\pi)\right)\,,
\eeq
where $\mathcal{H}=T^{\tau}{}_{\tau}$ is the boundary energy density coming from the boundary stress tensor, the first integral is taken over the entire spacetime, and the integrals are performed modulo the Hamiltonian and momentum constraints.\footnote{Note that the Hamiltonian constraint does not imply that AdS energy vanishes. Rather the constraint implies that AdS energy is a boundary quantity, coming from the boundary stress tensor.} Fixing the boundary energy here amounts to inserting a delta function $\delta\!\left( E-\frac{1}{\beta} \int d^dx \sqrt{g^{(0)}}\,\mathcal{H}\right)$, and following the same argument as in the last Subsection this formally produces $\rho(E) e^{-\beta E}$, where the density of states may be expressed as a functional integral
\beq
	\rho(E)=\int [dg] [d\pi] \,\exp\!\left( i \int d^{d+1}x\,\pi^{ij}\partial_{\tau} g_{ij}\right)\delta\left( E-\int d^dx \sqrt{g^{(0)}} \,\mathcal{H}\right)\,,
\eeq
and we have scaled out $\beta$. This expression should only be understood as suggestive since the sum over metrics does not converge. Nevertheless we regard it as a good heuristic argument that this constraint really does fix energy in the dual description.

\subsection{Stabilizing wormholes}

Now let us turn back to the wormholes we reviewed in Section~\ref{sec:review}. Upon fixing a gauge, these wormholes arise as a one-parameter family of solutions to the gauge-fixed equations of motion. This parameter is a modulus, and the wormhole amplitude can be written as a moduli space integral over it. 

We also saw that these wormholes have the property that the two boundary energies are equal, $E_1 = E_2 = E$, and that $E$ is precisely the parameter that labels the wormhole. These wormholes are saddle points of the gravity action with respect to all other directions in field space besides the one-parameter family $E_1=E_2=E$, including in particular the orthogonal fluctuation of boundary energies $E_1 - E_2$. We would like to stabilize this modulus ``by hand'' using the constant energy constraint of the last Subsection, in such a way as to have a genuine, albeit constrained, saddle of Einstein gravity. 

We achieve this by constraining the total energy $E_1+E_2$. This can be done with a delta function constraint, or with a bit of Gaussian broadening. The reason why we choose  this constraint
is because the gradient of $E_1+E_2$
is precisely the direction in field space along which the wormhole action varies. We thus pin down the modulus direction, and the wormhole is stationary with respect to fluctuations in all other non-constrained directions, and so we have a saddle to this unconventional boundary value problem in Einstein gravity.

While we have found these wormholes in a particular gauge, the result -- that we find a wormhole saddle of the two-boundary problem once we introduce this non-local constraint -- is gauge-invariant, precisely because this boundary condition is gauge-invariant.

As a simple model for what is happening, let us write the wormhole amplitude as an integral over $E_1$ and $E_2$ with $E_1-E_2$ very small, 
\beq
	\int dE_1 dE_2 \,e^{-(\beta_1+\beta_2)\frac{E_1+E_2}{2} - \frac{\kappa\,(E_1-E_2)^2}{2}}\tilde{f}(E; \beta_1, \beta_2)\left( 1 + O(G)\right)\,,
\eeq
where as for the wormhole amplitude the integrand is only accessible in the semiclassical domain for sufficiently large energies. Constraining $E_1+E_2=2E$, so that $E_1 = E+\frac{\delta E}{2}$ and $E_2 = E-\frac{\delta E}{2}$, the integral over the remaining fluctuation $\delta E=E_1 - E_2$ may be performed by saddle-point approximation around the saddle $\delta E = 0$, i.e. around a configuration with $E_1 = E_2$.

If instead of fixing the direction in which the action varies, suppose we fix the energy $E_1$ on the first boundary, perhaps with some resolution. However this moves the minimum of the action in the unconstrained direction $E_2$ slightly, from $E_2 = E_1$ to $E_2\approx E_1 - \frac{\beta_1+\beta_2}{2\kappa}$. But, once the two energies are unequal, we are no longer dealing with the wormholes we started with, and the ensuing metric is a non-saddle with respect to fluctuations of the metric other than $E_2$. Similar statements apply if we fix a different combination of energies than $E_1+E_2$.

Because we are constraining the total boundary energy, the constrained wormhole contributes to the two-boundary partition function, sliced according to total energy.  The constrained wormhole is, in fact, a constrained saddle.  Writing the two-boundary partition function as an integral over both boundary energies,
\beq
	\left\langle \text{tr}\!\left( e^{-\beta_1 H}\right)\text{tr}\!\left( e^{-\beta_2 H}\right)\right\rangle= \int dE_1 dE_2 \,\langle \rho(E_1)\rho(E_2)\rangle \,e^{-\beta_1 E_1 - \beta_2 E_2}\,,
\eeq
where we are being agnostic as to whether there is some averaging or not, the wormhole is a saddle that contributes to
\beq
	\left\langle \text{tr}\!\left( e^{-\beta_1 H}\right)\text{tr}\!\left( e^{-\beta_2H}\right)\right\rangle_{\mathcal{C}}= \int dE_1 dE_2 \langle \rho(E_1)\rho(E_2)\rangle \,e^{-\beta_1 E_1-\beta_2E_2} \delta(E_1 + E_2 - 2E)\,.
\eeq

Here we find one of the main lessons of the present work. Even though the wormhole contributes to the two-boundary amplitude and so seemingly violates factorization, the quantity to which it is dual does not factorize on account of the non-local constraint on energies even if the boundary lacks an ensemble average. 

For a few reasons it is more practical to smear in total energy with a Gaussian window of some small width $\Delta$, rather than to exactly fix total energy. In other words, consider
\beq
	\int dE_1 dE_2 \,\langle \rho(E_1)\rho(E_2)\rangle\, e^{-\beta_1 E_1-\beta_2 E_2} \,e^{-\frac{(E_1+E_2-2E)^2}{2\Delta^2}}\,,
\eeq
which can be achieved in the bulk by broadening the constraint on total energy. The wormhole remains a saddle-point approximation to this unconventional problem in AdS gravity, but there are two further benefits. The first is that it is easier to deal with the contribution to this quantity from the disconnected geometry, namely two copies of a Euclidean black hole. The second is that, upon analytic continuation $\beta_1 = \beta+iT$ and $\beta_2=\beta-i T$, on the boundary we are dealing with a microcanonical version of the spectral form factor. It is microcanonical in the sense that the total energy has been fixed, and for many-body chaotic systems this has a nice form when that total energy is smeared over a window much larger than the mean level spacing $\sim e^{-S}$, but much less than the curvature of the density of states.  This form of smearing is apparently present in gravity.

Denoting the form factor as $\widetilde{Z}_{E,\Delta}(\beta;T)$, the gravitational result coming from the wormhole is, for $T\gg \beta$,
\beq
\label{E:smoothRamp}
	\widetilde{Z}_{E,\Delta}(\beta;T)\approx T e^{-2\beta E}\,,
\eeq
where the exponential factor is $e^{-S}$ with $S$ the constrained wormhole action, and we are including a factor of $T$ that arises from a gravitational zero mode that translates one boundary in time relative to the other as in~\cite{Saad:2018bqo}. Other fluctuations are expected to be suppressed in the large $T/\beta$ limit. This is a ramp, which as we have mentioned before is a universal prediction for chaotic systems. 

The ramp coming from this wormhole is smooth with time. This smoothness is automatic in many ensemble averages, like random matrix theories; without an ensemble, the form factor is instead an extremely erratic function of time, with a ramp only appearing upon a suitable rolling time average. See e.g.~\cite{prange1997spectral, Cotler:2016fpe, Cotler:2017jue} for further discussion. However, the fact that a smooth ramp can appear by averaging over time windows implies that it can also emerge by a smearing in the energy domain.
In gravity the mean level spacing of the black hole spectrum is non-perturbatively small in the gravitational coupling, of order $e^{-\frac{1}{G}}$, or in terms of a large $N$ parameter, of order $e^{-N^m}$ where $m=\frac{3}{2}, 2, 3$ depending on whether one is studying ABJM theory at small $k$, a large $N$ gauge theory, or the theory on coincident M5 branes. We stress that because this level spacing is non-perturbatively small, bulk effective field theory computations around known saddles cannot resolve it, as they produce power series approximations in powers of $G$ or equivalently in powers of $\frac{1}{N}$.  Thus bulk effective field theory naturally comes with a resolution scale, any function of $N$ that is smaller than a power series in $1/N$. This includes non-perturbatively suppressed scales that are still parametrically larger than the mean-level spacing, like $N^pe^{-N^m}$. Smearing over such an energy domain would lead to fluctuations in the form factor over non-perturbatively small frequencies, which is completely consistent with the smooth ramp in~\eqref{E:smoothRamp}. We also note that in these wormholes, because the difference of energies $E_1-E_2$ is a stable direction in the spectrum of fluctuations, large energy differences are suppressed.

Using bulk effective field theory we then cannot distinguish between such a smearing in energies and an ensemble average. Both lead to a smooth ramp to the precision accessible to a bulk effective field theorist. The difference between the two lies in non-perturbatively small and/or fast oscillations in the form factor, and so we cannot distinguish between these possibilities using current methods in the bulk. However within the AdS/CFT correspondence where we have strong priors for factorization and against averaging on the boundary, we have good reason to believe that gravity is smearing over energy windows rather than over couplings. By pushing for this interpretation of wormholes in AdS, we are effectively advocating for a ``have your cake and eat it too'' resolution of the factorization paradox, in that we are suggesting that the wormholes that are reliably accessible in bulk effective field theory describe sensible physics, like a ramp in the form factor, while being consistent with factorization.

Lastly let us comment about brane nucleation instabilities. We can uplift these 5d wormholes to asymptotically Euclidean AdS$_5\times\mathbb{S}^5$ wormholes in type IIB supergravity~\cite{Cotler:2021cqa}, and fixing the total energy renders them saddle points of the supergravity action. When there is a torus boundary, the ensuing wormholes are always unstable to the nucleation of D3-$\overline{\text{D}3}$ brane pairs, although this instability is ultimately due to the moduli space of vacua for flat space $\mathcal{N}=4$ supersymmetric Yang-Mills theory. There is reason to expect this instability to be lifted upon turning on small supersymmetry-breaking deformations that, with a single flat boundary, generate a stable potential on the moduli space. For $\mathbb{S}^1\times \mathbb{S}^3$ boundary and real $\beta_1,\beta_2$, wormholes with large energy above a threshold $E\approx 3.3 E_0$ are unstable to the nucleation of brane pairs, while those with energies between $E_0$ and this threshold are stable. However upon continuing $\beta_1$ and $\beta_2$ to $\beta \pm i T$, all of these wormholes are stable against nucleation for sufficiently late times $T \gg \beta$.

\section{Wormholes are at least as heavy as black holes}
\label{sec:heavier}

As we explained in the Introduction, we only expect wormholes to exist above the small black hole threshold. This inequality was emphasized recently by Schlenker and Witten~\cite{Schlenker:2022dyo} in their analysis of pure AdS$_3$ gravity, and they conjectured that, appropriately understood, the same conclusion should hold in higher dimensions.

Indeed, the wormholes we studied in the last Section all have energies above the corresponding black hole threshold, as noted in~\cite{Cotler:2021cqa}. However, that dataset is a bit limited. We have found analytic solutions for wormhole metrics in Euclidean AdS$_5$ with $\mathbb{S}^1\times \mathbb{S}^3$ boundary, and in general dimension with torus boundary, both without electric flux or angular momentum. In~\cite{cotler2020ads} where we computed the torus times interval amplitude of pure three-dimensional gravity, we also found wormholes with torus boundary that carried angular momentum, and in that setting the wormholes always have energy above the corresponding rotating BTZ threshold.

In this Section we establish a much more general result for wormholes in general dimension with $\mathbb{S}^1\times \mathbb{S}^{d-1}$ or $\mathbb{T}^d$ boundary, with or without electric flux, as long as the wormhole carries no angular momentum. In this wide arena we are unable to find analytic solutions for the wormhole metrics, but nevertheless we can show that their actions take the same form as what we reviewed above,
\beq
\label{E:wormholeAction}
	S = (\beta_1+\beta_2)E\,,
\eeq
where the boundary energies $E_1=E_2=E$ all are above the small black hole threshold $E_0$ for the relevant boundary topology and electric flux. We can then constrain the total energy as in the last Section and turn these wormholes into genuine, albeit non-standard saddles of Einstein-Maxwell theory. Informally, our result here is that the wormholes so obtained are at least as heavy as black holes. This is consistent with the outlook of Schlenker and Witten~\cite{Schlenker:2022dyo}.

To study these wormholes we follow the same algorithm from~\cite{Cotler:2021cqa} that we reviewed in Section~\ref{sec:review}. Namely, we study the gauge-fixed gravitational theory, with a radial gauge-fixing so that the line element is of the form
\beq
	ds^2 = d\rho^2 + g_{ij}(x,\rho)dx^i dx^j\,,
\eeq 
with $\rho$ the ``radial'' coordinate.  The Einstein-Maxwell equations of motion read
\begin{align}
\begin{split}
\label{E:constrainedEinstein}
	R_{ij} - \frac{1}{2}\,R\,g_{ij} + \Lambda \, g_{ij} &= 8 \pi G \left(F_{i}^{\,\,\,\alpha}F_{j \alpha} - \frac{1}{4} g_{ij} F^{\alpha\beta} F_{\alpha \beta}\right) \,, \\
\nabla^\nu F_{\mu \nu} &= 0\,.
\end{split}
\end{align} 
The Einstein's equations presented are the components that follow from variation with respect to the $ij$ components of the metric. The radial components of the metric have been gauge-fixed and so we need not solve their equations of motion. 

It is convenient to use a new radial coordinate $r = r(\rho)$.  Then our line elements are fixed to have the form $ds^2 = \rho'(r)^2\,dr^2 + g_{ij}( x^i,r) \, dx^i dx^j$.
Let us consider geometries with $\mathbb{S}^1\times\mathbb{S}^{d-1}$ boundary that preserve translations along the $\mathbb{S}^1$, rotations along the $\mathbb{S}^{d-1}$, and parity. The most general line element and Maxwell field satisfying these conditions is
\begin{align}
	ds^2 =  f_1(r) \,d\tau^2 + f_2(r) \, dr^2 + r^2 d\Omega_{d-1}^2\,, \qquad A_\mu \,dx^\mu = A_{\tau}(r)\, d\tau\,,
\end{align}
where we have used the freedom to choose a new radial coordinate to pick the sphere warp factor to be $r^2$. 
There is a conserved electric flux
\begin{equation}
	-iQ = \sqrt{g} g^{\tau\tau} g^{rr} A_{\tau}'(r) \qquad \Longrightarrow \qquad A_{\tau}'(r) =-i Q \frac{\sqrt{f_1(r)f_2(r)}}{r^{d-1}} \,,
\end{equation}
corresponding to a real flux in Lorentzian time $t=-i\tau$.
The $\tau\tau$ component of Einstein's equations then becomes a conservation equation for the $rr$ warp factor,
\begin{equation}
\label{E:grrconserve1}
	\frac{d}{dr}\left[r^{d-2}\left(\frac{1}{f_2(r)} - \left(r^2 + 1- \frac{8 \pi G}{(d-2)(d-1)} \,\frac{Q^2}{r^{2(d-1)}}\right)\right)\right] = 0\,.
\end{equation}
Note that the $\tau \tau$ warp factor decouples. 
The most general solution is
\begin{equation}
\label{E:warpfactorf1}
f_2(r) = \left(r^2+1 - \frac{m}{r^{d-2}}+ \frac{q^2}{r^{2(d-1)}}\right)^{-1}\,,
\end{equation}
with $Q^2 = \frac{(d-2)(d-1) \,q^2}{8 \pi G}$ and $m$ an integration constant. 

So far we are on our way to constructing either the standard charged Euclidean AdS black hole, or a wormhole. To find the black hole we solve the remaining Einstein's equations, and in particular the $rr$ component implies $f_1(r)=1/f_2(r)$. However we wish to find a wormhole solution to~\eqref{E:constrainedEinstein}, for which we need not solve the $rr$ component of Einstein's equations since we have gauge-fixed $g_{rr}$. Having solved the Maxwell's equations and the $\tau\tau$ component of the Einstein's equations, we need only solve the angular components to obtain the $\tau \tau$ warpfactor $f_1(r)$. This cannot be done in closed form except for $q=0$ in $d=2,4$, however as we will see, we do not need to solve it in order to find the wormhole action~\eqref{E:wormholeAction} or the lower bound on boundary energy.

It is worth detouring a little bit to discuss the black hole. In that case the parameter $m$ is essentially the black hole mass, and the black hole only exists for sufficiently large mass such that 1/$f_2(r)$ has a real root. Let us denote the outermost real root as $r=r_+$, the horizon in the Lorentzian domain, and also denote the lower bound on $m$ as $m_0(q)$. For $q=0$ the lower bound is simply $m_0= 0$, while $m_0$ takes on a more complicated form at nonzero charge. For $m>m_0(q)$ there is a coordinate singularity at $r=r_+$; the geometry is manifestly smooth in terms of the coordinate $\varepsilon$ with $r=r_++\varepsilon^2$, with the $\mathbb{S}^1$ shrinking smoothly at $\varepsilon=0$ provided it has a particular periodicity. For $m=m_0$, one has an extremal black hole, while for $m<m_0$ the radial warp factor $1/f_2(r)$ has no real root and the geometry has a naked singularity at $r=0$.

Now we turn to wormholes, which obey a similar classification. If $m> m_0$ there are three possible outcomes. If $f_1(r)$ is nonzero for $r\geq r_+$, in particular at $r=r_+$, then we have half of a wormhole geometry. If $f_1(r)$ has a root $\tilde{r}$ for some $r>r_+$, then the circle is shrinking there and if it does so smoothly, we have an off-shell Euclidean black hole. Finally if $f_1(r)$ has a root at $r_+$ so that the circle shrinks smoothly there, then we again have a Euclidean black hole.

In contrast, if $m<m_0$, the only smooth geometry we can consider is one for which $f_1(r)$ shrinks smoothly at some $r>0$, in which case, again, we find an off-shell Euclidean black hole.

It would be surprising to find off-shell black holes of this sort, for either $m>m_0$ or $m<m_0$. Indeed, we do not have a proof that such configurations do not exist, but we have made some numerical searches for these geometries and have not found examples with asymptotically Euclidean AdS regions.

So we have seen that our geometries describe half of a wormhole as long as $m$ is above the black hole threshold $m_0(q)$. To find the other half we need to glue two of these geometries together. A priori the two halves are characterized by different charges and masses, $(q,m)$ in one half, and $(q',m')$ in the other. However as we have seen, Maxwell's equations and the $\tau\tau$ component of Einstein's equations take the form of conservation equations, guaranteeing that $m=m'$ and $q=q'$. So the two halves are characterized by the same $f_2(r)$. These can be smoothly glued together at $r=r_+$.

Given a wormhole of fixed charge $q$ and mass $m$, one can find the remaining warpfactor $f_1(r)$ by shooting from the coordinate singularity $r=r_+$ outward to the two boundaries. This can be achieved by passing to a covering coordinate $r=r_+\cosh(u)$ as for the black hole, where $u$ is valued on the real line with the two boundaries attained as $u \to \pm \infty$. One can shoot from $u=0$ provided that one starts with a solution to the angular component of Einstein's equations there. The most general such solution has two parameters,
\beq
	f_1(u) = c_1 + c_2 u + O(u^2)\,,
\eeq
and these two parameters map to the sizes $\beta_1$ and $\beta_2$ of the boundary circles attained as $u\to \pm \infty$. 

In summary, for fixed charge $q$ and mass $m$ above the black hole threshold $m_0(q)$, we find a two-parameter family of wormholes labeled by the sizes of the boundary circles $\beta_1$ and $\beta_2$.

The fact that the wormhole has the same ``mass'' $m$ in the two asymptotic regions suggests that the boundary energies coincide as claimed, since boundary energy is the mass up to counterterms introduced in holographic renormalization. Recall that the boundary stress tensor on the boundary achieved as $r\to \infty$ is 
\beq
	T^{\mu\nu} =\lim_{r \to \infty} \left[  \frac{\sqrt{\gamma}}{8\pi G\sqrt{g^{(0)}}}\left( K^{\mu\nu} - K \gamma^{\mu\nu}\right) + T^{\mu\nu}_{\rm CT}\right]\,,
\eeq
where we work with cutoff slices at fixed $r$ which have an induced metric $\gamma_{\mu\nu}$ and extrinsic curvature $K_{\mu\nu}$, the boundary metric is $g^{(0)}_{\mu\nu}$, and $T^{\mu\nu}_{\rm CT}$ is the contribution to the stress tensor from counterterms on the cutoff slice that arise from holographic renormalization. The latter are a function only of the boundary metric, and so are identical for a black hole and for a wormhole.

Because the precise form of the counterterm stress tensor $T^{\mu\nu}_{\rm CT}$ is somewhat complicated and depends on the dimension, to compare boundary energies of the two wormhole halves with each other and with the boundary energy of a black hole, it is convenient to consider the difference between the boundary energy of a wormhole with that of a black hole having the same parameters $m$ and $q$. To do so we require the asymptotic behavior of the warp factor $f_1(r)$ near one asymptotic boundary, which can be obtained by solving the angular components of Einstein's equations at large $r$. Suppose we are close to one of the two asymptotic boundaries where the thermal circle has a size $\beta_1$. Then with $\tau \sim \tau + 1$ we have the result
\beq
	f_1(r) = \beta_1^2\left( r^2 + 1 -\frac{\tilde{m}}{r^{d-2}} + O(r^{-d+1})\right)\,,
\eeq
for a constant $\tilde{m}$ that we cannot determine without constructing the full two-sided geometry. Fortunately the boundary energy is completely insensitive to that constant, and we have
\beq
	T^{\tau}{}_{\tau, \, \text{wormhole}} - T^{\tau}{}_{\tau,\, \text{black hole}} = 0\,,
\eeq
so that the boundary energy densities for the wormhole and black hole coincide. Boundary energy is just the integral of this density over the boundary sphere, so we see that for fixed mass $m$ and charge $q$, the energies on the two boundaries of the wormhole coincide with that of the black hole with the same mass and charge, and so with each other. In particular we have $E_1=E_2=E\geq E_0(q)$, where $E_0(q)$ is the black hole threshold at charge $q$.

Because boundary energy is conjugate to the circle size $\beta$, this result also allows us to reconstruct the wormhole action $S_{\rm wormhole}(\beta_1,\beta_2)$. These geometries are invariant under translations along the $\tau$ circle, which implies $S_{\rm wormhole}(\lambda \beta_1,\lambda \beta_2)=\lambda \,S_{\rm wormhole}(\beta_1,\beta_2)$ and so $S_{\rm wormhole}(\beta_1,\beta_2) = \beta_1 F(\beta_1/\beta_2)$. But in the canonical ensemble where we fix the boundary charges to be equal,
\beq
	E_1 = \frac{\partial S_{\rm wormhole}}{\partial \beta_1}= \frac{\partial S_{\rm wormhole}}{\partial \beta_2} = E_2\,,
\eeq
which has a unique solution
\beq
	S_{\rm wormhole} = (\beta_1 + \beta_2)E\,,
\eeq
up to an integration constant which is independent of the boundary data. This result can be verified explicitly in examples; for instance, a particularly nice case is a 5d wormhole carrying a perturbatively small amount of flux.

As in our discussion in Section~\ref{sec:fixing}, we may turn these wormholes into bona fide saddles by constraining the total energy $E_1+E_2$. Fixing the charge $q$ as well, we see that, as advertised, these wormholes always have an energy above the small black hole threshold $E_0(q)$. 

So far we have considered wormholes with $\mathbb{S}^1\times \mathbb{S}^{d-1}$ boundary. However, the analysis for wormholes with torus boundary goes through \emph{mutatis mutandis}. The starting point is a line element and Maxwell field
\beq
	ds^2 = f_1(r)^2d\tau^2 + r^2 d\vec{x}^2 + f_2(r)^2 dr^2\,, \qquad A_{\mu}dx^{\mu} = A_{\tau}(r)d\tau\,,
\eeq
where $\vec{x}$ parameterizes a spatial torus. The Maxwell field is fixed up to an integration constant in terms of the electric flux, and the $\tau\tau$ component of Einstein's equations becomes a conservation equation for the radial warp factor $f_2(r)$ with the solution
\beq
	f_2(r) = r^2 - \frac{m}{r^{d-1}} + \frac{q^2}{r^{2(d-1)}}\,.
\eeq
The rest of the argument goes through similarly.

\section{Parametric correlations in chaotic quantum systems and AdS/CFT}
\label{sec:gcorr}

\subsection{Parametric correlations and RMT}

A celebrated result in quantum chaos and random matrix theory is the universality of short-range level-level statistics of energy eigenvalues, which can be packaged into the connected part of the spectral form factor.  Originally computed in Gaussian matrix models by Dyson and others in the 1960's~\cite{dyson1962statisticalone, dyson1962statisticaltwo, dyson1962statisticalthree, mehta2004random}, these same level-level statistics were quantitatively recapitulated in a large variety of few-body and ultimately many-body quantum systems through combinations of numerics and ingenious analytics, making a strong case for universality (for reviews see~\cite{guhr1998random, haake2013quantum}).  In the 1980's and 1990's, it was shown that a large class of single-trace matrix models robustly possess these same level-level statistics~\cite{Ambjorn:1990wg, brezin1993universality}, and a universality argument for a much larger class of systems was given by Efetov using his SUSY sigma model approach~\cite{efetov1983supersymmetry, efetov1999supersymmetry}.  Recently, this approach was given a more robust understanding in the language of symmetry breaking~\cite{Altland:2020ccq}, better motivating its application to general quantum chaotic systems.  
It has also been shown to reproduce short-range level-level statistics of certain quantum many-body systems in~\cite{Altland:2017eao, Altland:2020ccq, Altland:2021rqn}.

The above results entail comparing pairs of energies within a single Hamiltonian, perhaps with some disorder average.  This encompasses most of the standard paradigm of random matrix theory, and its connection to chaotic quantum systems.  A conceptually different set of questions arises if we were to compare energies between two distinct Hamiltonians.  For instance, suppose we have a quantum many-body system with Hamiltonian $H(g)$ which depends on a coupling $g$.  We might ask if there are correlations between the eigenvalues of $H(g)$ and $H(g + \delta g)$, particularly when $\delta g$ is small.  This is a natural arena for study in physical systems, namely how the statistics of their spectra depend on coupling constants.

This set of questions can be fruitfully studied in toy random matrix models, for instance comparing the spectra of $H_0 + g \, H_1$ and $H_0 + (g + \delta g) H_1$ where $H_0, H_1$ are Gaussian random matrices~\cite{d1995universal}.  The broader study of such correlations is called parametric random matrix theory on account of the parametric dependence of the coupling $g$.  In this context, it is natural to study an analogue of the connected spectral form factor, in which $H(g)$ and $H(g + \delta g)$ are compared.  In a number of analytic and numerical examples, it was found that there is a common answer which appeared to be universal, independent of most details of the systems (see~\cite{guhr1998random} for a review).  This was corroborated by Simons and Altshuler using the Efetov SUSY sigma model approach~\cite{Simons:1993zza, simons1993universalities}.   Taken together, this body of numerical and analytic results strongly suggests the universality of certain level-level correlations between two distinct quantum systems, even if those systems are not disordered.  Here we review these particular correlations and their universality, and subsequently explain their role in quantum gravity.

\subsubsection{Existing results}

Consider a Hamiltonian $H(g)$ that depends on a coupling $g$. We begin by defining an auxiliary quantity $x$.  Let $E_n(g)$ be the $n$th energy level of $H(g)$, and suppose that $\mathcal{N}(\bar{E},\Delta)$ is the number of energy levels $E_n(g)$ in the window $[\bar{E} - \Delta, \bar{E} + \Delta]$.  We suppose that $\Delta$ is much larger than the mean level spacing, but is small relative to the curvature of the coarse-grained density of states and that the average level spacing does not change much in the window. Moreover, we ``unfold'' the energy spectrum so that within the window the mean level spacing is $1$. The unfolded energies are $\varepsilon_n = \overline{\rho} \,E_n$ where $\overline{\rho}$ denotes the coarse-grained density of states within the window. Then we define
\begin{equation}
\label{E:xdef1}
	x^2 \equiv \frac{1}{\mathcal{N}(\bar{E},\Delta)}\sum_{E_n \in [\bar{E} - \Delta, \bar{E} + \Delta]} \left(\frac{d\varepsilon_n}{dg}\right)^2 dg^2 =\overline{\left(\frac{d\varepsilon_n}{dg}\right)^2}\delta g^2\,,
\end{equation}
where $\delta g$ is a variation in the coupling $g$, and the bar denotes a smeared microcanonical average over the window. It will be important that we consider Hamiltonians for which, at a coupling $g$,
\begin{equation}
\label{E:dontChangeDOS}
\frac{1}{\mathcal{N}(\bar{E},\Delta)}\sum_{E_n \in [\bar{E} - \Delta, \bar{E} + \Delta]} \frac{d\varepsilon_n}{dg} =\overline{\frac{d\varepsilon_n}{dg}}= 0\,,
\end{equation}
which means that a small change in coupling $g\to g+\delta g$ does not change the mean density of states to linear order in $\delta g$. The quantity $x$ can then be thought of as a root-mean-square velocity of the eigenvalues when changing the coupling $g$. Note that $x$ is a one-replica quantity.

Consider the parametric density-density autocorrelator
\begin{equation}
\label{E:ddcorrelator1}
	k(\bar{E},\delta E;g,\delta g) =  \overline{ \delta\left( \bar{E}+\delta E-E_n\left(g+\frac{\delta g}{2}\right)\right)\delta\left(\bar{E}-E_m\left(g-\frac{\delta g}{2}\right)\right)}-\bar{\rho}^2\,.
\end{equation}
This is a two-replica quantity, where we have one copy of the spectrum at coupling $g+\frac{\delta g}{2}$ and the other at coupling $g-\frac{\delta g}{2}$.  Assuming that our Hamiltonian does not have time-reversal symmetry (in particular, it is GUE class), we have the universal result~\cite{Simons:1993zza, simons1993universalities} that $k$ only depends on the change in coupling $\delta g$ through the quantity $x$ as
\begin{equation}
	k(\bar{E},\delta E,x) = \frac{\bar{\rho}^2}{2}\int_{-1}^1 d\lambda \int_1^\infty d\lambda_1 \, e^{- \frac{\pi^2 x^2}{2} (\lambda_1^2 - \lambda^2)} \cos\left( \pi \,\bar{\rho}\,\delta E\, (\lambda - \lambda_1)\right)\,,
\end{equation}
which is valid for $\bar{\rho}\, \delta E = \delta \varepsilon \sim O(1)$ and $x^2 \sim O(\bar{\rho})$.
This formula has been corroborated by numerical experiments and tractable analytic examples (see~\cite{guhr1998random, wilkinson1999parametric} for overviews).  In these examples, either the theories in question are disorder averaged, or otherwise the spectral correlations are considered between two individual Hamiltonians and the answer is smoothed out by smearing over an energy window.  In this manner, disorder averaging and energy smearing are complementary approaches, and for certain averaging and smearing schemes the leading final result should be agnostic. The quantity $\bar{\rho}\,\delta E$ is the unfolded energy difference. To obtain the microcanonical spectral form factor in the band we take the Fourier transform with respect to $\delta E$ and perform the residual $\lambda$ and $\lambda_1$ integrals to find
\begin{align}
\begin{split}
	Z_{\text{micro}}(\bar{E},t; x) & =\overline{\text{tr}\left( e^{-i h(g+\delta g/2)t}\right)\text{tr}\left( e^{i h(g-\delta g/2)t}\right)}
	\\
	& =  \frac{\bar{\rho}}{2\pi\, t x^2}\left( 2 \sinh\!\left(\frac{1}{2}(2\pi - t) t x^2\right) \, \Theta(t - 2\pi) +2\,e^{- \pi t x^2} \sinh\!\left(\frac{t^2 x^2}{2}\right) \right)\,,
\end{split}
\end{align}
where $h(g)$ is the unfolded Hamiltonian at coupling $g$, $t=\frac{T}{\bar{\rho}}$ is the unfolded time conjugate to $\bar{\rho}\,\delta E$ (with $T$ the ordinary time conjugate to $\delta E$), and $\Theta(t)$ is the Heaviside step function. This formula is valid for $t$ parametrically larger than $\bar{\rho}^{-2}$.  To understand this result, we first take the $x \to 0$ limit to arrive at the famous ramp and plateau
\begin{equation}
\label{E:famousramp}
Z_{\text{micro}}(\bar{E},t; 0) = \frac{\bar{\rho}}{2\pi } \left(t + (2\pi - t)\,\Theta(t- 2\pi) \right)\,,
\end{equation}
where before time $t=2\pi$ we have a linear ramp with slope $2\pi$, and after that we have a plateau. The transition takes place at unfolded time $t = 2\pi$, i.e.~a real time $T = 2\pi \bar{\rho}$, due to our unfolding and rescaling of the energy spectrum.

The limit we will study is that of small  $t$ with $x^2 t$ fixed, for which
\begin{equation}
\label{E:parametricExpectation}
	Z_{\text{micro}}(\bar{E},t; x) \approx \frac{\bar{\rho}\,t}{2\pi }\, e^{- \pi  x^2 t} = \frac{T}{2\pi}e^{-\pi X^2T} \,,
\end{equation}
where we have defined $X^2 = \frac{x^2}{\bar{\rho}}$. Here we see that the linear ramp in~\eqref{E:famousramp} is squelched by the exponential decay.  Physically, this is because the two Hamiltonians $H(g\pm \delta g/2)$ in the form factor are at different couplings; as the difference between the couplings increases, the mutual long-range level correlations between the Hamiltonians become exponentially decorrelated.

Another way of understanding this decorrelation is to go back to energy space.  In the $x \to 0$ limit, $k(\bar{E},\delta E; x\to 0) = -\sin^2(\pi \,\bar{\rho}\,\delta E)/(\pi\,\bar{\rho}\,\delta E)^2$ which is the negative square of the Dyson sine kernel, i.e.~the usual answer in standard RMT.  For large energy separations this has power-law decay $\sim - 1/\delta E^2$ whose Fourier transform to real-time is responsible for the ramp in the spectral form factor.  Turning on $x$, this power law gets augmented to be $\sim - 1/(\delta E^2 + \pi^2 x^4\bar{\rho}^{-2})$, and so the shifted pole yields an exponential decay upon Fourier-transforming to real-time.

To summarize, the result from a variety of analytic and numerical examples is that, with either smearing or ensemble averaging, the microcanonical form factor with two replicas at slightly different couplings is smooth and decays exponentially. In real time $T = \bar{\rho}\,t$, the rate is $\pi X^2$, a one-replica quantity~\eqref{E:xdef1} that depends on the details of the Hamiltonian. 

\subsubsection{Evaluating the decay rate}

We would like to evaluate the rate $X^2=\frac{x^2}{\bar{\rho}}$ for theories with a holographic dual. So far we have written $x^2$ as a microcanonical average. However we could evaluate $x^2$ with any distribution we like, provided it is tightly enough peaked around the desired average energy $\bar{E}$. In particular we can do so with a thermal distribution.

Let us first return to $\frac{x^2}{\bar{\rho}}$ as a microcanonical average. Using that for $H(g+\delta g) = H(g) + O \delta g + O(\delta g^2)$ we have $\frac{d\varepsilon_n}{dg} = \bar{\rho}\,\frac{dE_n}{dg} = \bar{\rho} \,\langle E_n(g) |O|E_n(g)\rangle$. We then write the average as an integral
\beq
	X^2 = \frac{x^2}{\bar{\rho}} = \frac{\bar{\rho}}{\mathcal{N}} \int_{\bar{E}-\Delta}^{\bar{E}+\Delta} dE\,\rho(E) |\langle E|O|E\rangle|^2 = \int dE\,\Omega(\bar{E},E;\Delta) \,\rho(E)|\langle E|O|E\rangle|^2\,,
\eeq
where
\beq
	\Omega(\bar{E},E;\Delta) = \begin{cases} \frac{1}{2\Delta}\,, & E\in [\bar{E}-\Delta,\bar{E}+\Delta]\,, \\ 0 \,, & \text{otherwise}\,,\end{cases}
\eeq
is the relevant microcanonical distribution.  Now let us evaluate the average with a thermal distribution
\beq
	\frac{\rho(E)\,e^{-\beta E}}{Z(\beta)}\,,
\eeq
tuned so as to have a maximum value at $\bar{E}$. Because we have in mind many-body systems with a large number of degrees of freedom $\sim N$, this distribution has a very narrow width and the microcanonical and thermal averages will agree in the large $N$ limit. We then have
\begin{equation}
\label{E:xdef2}
X^2 = \int dE \,\frac{e^{- \beta E}}{Z(\beta)} \,\rho(E)^2\,|\langle E|O|E\rangle|^2\, \delta g^2\,.
\end{equation}
Written this way the constraint~\eqref{E:dontChangeDOS} reads
\begin{equation}
\label{E:dontChangeDOS2}
\int dE \,\frac{e^{- \beta E}}{Z(\beta)} \,\rho(E)\langle E|O|E\rangle = \langle O\rangle_{\beta} = 0\,,
\end{equation}
where the angled brackets denote a thermal average. This implies in particular that $\langle H(g+\delta g)\rangle_{\beta} = \langle H(g)\rangle_{\beta} + \delta g \langle O\rangle_{\beta} + O(\delta g^2) = \bar{E} + O(\delta g^2)$, i.e. the coupling does not shift thermal averages at the level of linear response, and in particular does not shift the mean level spacing.

We now write
\begin{align}
\begin{split}
\label{E:X2eq1}
	X^2 &= \int dE \int dE'\,\frac{e^{- \beta E}}{Z(\beta)} \,\rho(E)\,\rho(E')\,\langle E|O|E'\rangle\langle E'|O|E\rangle\, \delta( E-E')\, \delta g^2 
	 \\
	&=  \int \frac{dt}{2\pi} \int dE \int dE'\,\frac{e^{- \beta E}}{Z(\beta)} \,\rho(E)\,\rho(E')\,\langle E|O|E'\rangle\langle E'|O|E\rangle\, e^{it(E-E')}\, \delta g^2
	 \\
	&= \frac{1}{2\pi} \int dt \, \text{tr}\!\left(\frac{e^{- \beta H}}{Z(\beta)}\, O(t) O(0) \right) \delta g^2
	\\
	&= \frac{1}{2\pi} \int dt \,\langle O(t) O(0)\rangle_\beta\delta g^2=\frac{1}{2\pi} \int dt\, \frac{\langle \{O(t),O(0)\}\rangle_{\beta}}{2}\delta g^2\,,
\end{split}
\end{align}
and so we see that $X^2$ is ($\frac{1}{2\pi}\delta g^2$ times) the zero-frequency symmetrized thermal 2-point function of the perturbation $O$. This response function may be easily evaluated for theories with a holographic dual from a Euclidean black hole. 

In terms of the quantity $X$ the microcanonical form factor is expected to behave as
\beq
\label{E:parametricExpectation2}
	\overline{\text{tr}\left( e^{-iH(g+\delta g/2)T}\right)\text{tr}\left( e^{iH(g-\delta g/2)T}\right)} \approx \frac{T}{2\pi}\,e^{-\pi X^2 T}\,,
\eeq
for $T$ large enough, but not too large so that it obeys $T \lesssim \bar{\rho}^{1/2}$. In the next Subsection we will compute this form factor and find that it decays exponentially at the correct rate for holographic theories.

\subsection{Correlations between holographic theories at different couplings}

We turn now to the study of holographic theories. We use the methods of Section~\ref{sec:fixing}, in particular we will constrain the two-boundary amplitude in order to find a wormhole saddle. To compare with the expectation~\eqref{E:parametricExpectation} for chaotic systems we must independently compute $X^2$ via~\eqref{E:X2eq1}. Remarkably we find a match between the two for all two-derivative bulk effective theories obeying the assumption~\eqref{E:dontChangeDOS2} upon which the expectation~\eqref{E:parametricExpectation} rests.

That assumption, that the perturbing operator $O$ has vanishing thermal one-point function $\langle O\rangle_{\beta} = 0$, is obeyed for two-derivative effective theories where the black hole solution has scalar fields sitting at constant values, i.e. at an extremum of the scalar potential. Higher-derivative corrections like $\int d^{d+1}x\sqrt{g}\, \phi W_{\mu\nu\rho\sigma}W^{\mu\nu\rho\sigma}$~\cite{Meltzer:2017rtf} generate a small but nonzero thermal one-point function, but these effects can be approximately ignored for this analysis. 

We begin by relating $X^2$~\eqref{E:X2eq1} to a left-right correlator obtained from a two-sided black hole. As has been long known, black holes in AdS are dual to thermal states of the dual theory where the Hawking temperature of the black hole equals the temperature of the dual theory. Eternal two-sided black hole can be regarded as the dual of the thermofield double state, or, more conveniently for us, the dual to a Keldysh closed-time-path contour where we consider 
\beq
	Z_{\rm SK}[A_1,A_2]=\text{tr}\left( \mathcal{U}[A_1]e^{-\beta H}\mathcal{U}[A_2]^{\dagger}\right)\,,
\eeq
where $A_1$ and $A_2$ represent independent sources and $\mathcal{U}[A_1]$ is the infinite-time evolution operator in the presence of the source. By varying with respect to the sources and using that the path integral that computes $Z_{\rm SK}$ has the appropriate contour time ordering, this partition function encodes thermal two-point functions with any time-ordering, i.e.~retarded, advanced, and symmetrized, while all thermal four-point functions may be obtained from a closed-time-path contour with four real-time segments, etc. In the holographic context we may think of the $1$ sources as living on one asymptotically AdS boundary of the two-sided black hole, and the $2$ sources as living on the other.\footnote{With natural boundary conditions, the two-sided black hole computes a closely related partition function $\text{tr}\left( e^{-\frac{\beta H}{2}}\mathcal{U}[A_1]e^{-\frac{\beta H}{2}}\mathcal{U}[A_2]^{\dagger}\right)$~\cite{Maldacena:2001kr, Herzog:2002pc}. However, the zero-frequency two-point function of an operator $\mathcal{O}$ is equally well-encoded in this partition function as in $Z_{\rm SK}$.} We have
\beq
	\delta \ln Z_{\rm SK} = i \int d^dx_1 \sqrt{g_1} \,\mathcal{O}(x_1)\delta A_1(x_1) -i \int d^dx_2 \sqrt{g_2}\,\mathcal{O}(x_2)\delta A_2(x_2)\,,
\eeq
where $A$ is conjugate to the operator $\mathcal{O}$ and we are notating positions on the two real-time segments of the contour as $x_1$ and $x_2$.

In thermal physics it is convenient to pass from the $1, 2$ basis to the ``$r, a$'' basis of average and difference fields, 
\begin{align}
\begin{split}
	A_r(x) &= \frac{1}{2}(A_1(x)+A_2(x))\,, \qquad A_a(x) = A_1(x)-A_2(x)\,,
	\\
	\mathcal{O}_r(x) &= \frac{1}{2}(\mathcal{O}_1(x)+\mathcal{O}_2(x))\,, \qquad \mathcal{O}_a(x) = \mathcal{O}_1(x)-\mathcal{O}_2(x)\,,
\end{split}
\end{align}
and when the two metrics on the two contours are identical $g_1=g_2$ we have
\beq
	\delta \ln Z_{\rm SK} = i\int d^dx \sqrt{g}\left( \mathcal{O}_r(x) \delta A_a(x) + \mathcal{O}_a(x) \delta A_r(x)\right)\,.
\eeq
By varying twice with respect to the sources we obtain a $2\times 2$ matrix of two-point functions, which with a little algebra can be written as
\begin{align}
\begin{split}
	G_{ra}(1,2) &=\frac{\delta^2\ln Z_{\rm SK}}{\delta A_a(1)\delta A_r(2)} =-\Theta(t_1-t_2) \langle [\mathcal{O}(1),\mathcal{O}(2)] \rangle_{\beta}\,,
	\\
	G_{ar}(1,2) & = \frac{\delta^2\ln Z_{\rm SK}}{\delta A_r(1)\delta A_a(2)} = \Theta(t_2-t_1) \langle[\mathcal{O}(1),\mathcal{O}(2)]\rangle_{\beta}\,,
	\\
	G_{rr}(1,2) & = \frac{\delta^2\ln Z_{\rm SK}}{\delta A_a(1)\delta A_a(2)} = -\frac{1}{2}\langle\left\{ \mathcal{O}(1),\mathcal{O}(2)\right\}\rangle_{\beta}\,,
	\\
	G_{aa}(1,2) & = \frac{\delta^2 \ln Z_{\rm SK}}{\delta A_r(1)\delta A_r(2)} = 0\,,
\end{split}
\end{align}
where $\langle A\rangle_{\beta} = \text{tr}\left( \rho A\right)$ for $\rho = \frac{e^{-\beta H}}{\text{tr}(e^{-\beta H})}$ the thermal density matrix, we implicitly set $a$-type sources to vanish after taking the derivatives, and operators are evolved in the presence of the $r$-type sources. We are also denoting the time and space of the first and second insertions collectively as ``1'' and ``2.'' These response functions are (proportional to) the retarded, advanced, and symmetrized two-point functions respectively. By the by, for thermal response functions, the fact that $\rho$ is the translation operator in imaginary time implies an additional symmetry, often called the KMS condition, which relates the response functions as
\beq
	G_{rr}(\omega;x_1,x_2) = \frac{1}{2}\coth\left( \frac{\beta \omega}{2}\right)\left (G_{ra}(\omega;x_1,x_2)-G_{ar}(\omega;x_1,x_2)\right)\,,
\eeq
where we have Fourier transformed in time and $x_1,x_2$ denote the spatial positions of the two insertions. This  is the fluctuation-dissipation theorem.

Now, the zero-frequency function that computes $X^2$ in~\eqref{E:X2eq1} is proportional to the zero-frequency symmetrized function of $O$,
\beq
\label{E:simpleX2}
	X^2 =-\frac{1}{2\pi} G_{rr}(\omega=0) \delta g^2\,.
\eeq
From its definition $G_{rr}(\omega=0)$ is negative-definite, giving $X^2\geq 0$ and so $e^{- \pi X^2 T}$ is an exponential decay as expected. Note also that by the fluctuation-dissipation theorem this corresponds to an $O(\omega)$ term in the retarded function as $\omega\to 0$. We have in mind couplings $g$ for local operators $\mathcal{O}$ through a deformation $\delta H = \int d^{d-1}x \sqrt{g}\,\mathcal{O}(x) \delta g(x)$ with $\delta g(x)$ constant, i.e. $O = \int d^{d-1}x \sqrt{g(\vec{x})}\mathcal{O}(\vec{x})$. We are abusing notation a bit here, in that we are letting $g$ do double duty, denoting a coupling constant as well as the spatial metric either on a torus or a sphere, depending on whether we are studying thermal physics on a torus or sphere. In any case the decay of the form factor is 
\beq
	\pi X^2 T=- \frac{1}{2} \int d^{d-1}x_1 d^{d-1}x_2 \,\sqrt{g(x_1)} \sqrt{g(x_2)}\, \frac{\delta^2 \ln Z_{\rm SK}}{\delta g_a(\omega=0,x_1)\delta g_a(\omega=0,x_2)}\delta g^2 T\,,
\eeq
where $x_1$ and $x_2$ denote spatial positions respectively. 

In AdS/CFT $ Z_{\rm SK}$ is the partition function of the two-sided black hole $Z_{\rm BH}$. $X^2$ is the Fourier zero mode of the cross-horizon two-point function of $\mathcal{O}$ across the black hole, and so also of its orbifold $t\sim t+T$ that produces the corresponding double cone. However, we cannot yet understand $X^2$ as a correlation function obtained from the double cone. The obstruction is that the double cone has tree-level moduli, most notably the energy perceived on the two boundaries. In order for $X^2$ to be realized as a two-point function across the double cone, we must somehow fix the double cone moduli, so that they are the same values as for the black hole we started with. That is, we must fix boundary energy. Let us return to that in a moment. Having done so, we may then interpret the time integral that produces the zero frequency function and the factor of $T$ as integrals over the two times on the boundary of the double cone, giving
\begin{align}
\begin{split}
	\pi X^2 T &= - \frac{1}{2} \int dt_1 d^{d-1}x_1 dt_2d^{d-1} x_2 \sqrt{g(x_1)}\sqrt{g(x_2)} \,\frac{\delta^2 \ln Z_{\rm DC}}{\delta g_a(t_1,x_1)\delta g_a(t_2,x_2)} \delta g^2 
	\\
	&= - \frac{1}{2}\left.\frac{\partial^2 \ln Z_{\rm DC}}{\partial g_a^2}\right|_{g_a=0} \delta g^2\,.
\end{split}
\end{align}
In passing to the second line we are observing that we can rewrite the first in terms of the double cone where the coupling on the first boundary is $g+\frac{\delta g}{2}$ and the coupling on the second boundary is $g-\frac{\delta g}{2}$ so that the difference in couplings is a constant $a$-type source $g_a=g_1-g_2 = \delta g$. But this is nothing more than the leading term in the expansion of $\ln Z_{\rm DC}$ for small difference in coupling $\delta g$, i.e.
\beq
\label{E:decayOfDC}
	Z_{\rm DC} \propto e^{- \pi X^2 T + O(\delta g^3)}\,,
\eeq
where the linear term vanishes on account of $\langle O\rangle_{\beta} = 0$. 

This shift in $\ln Z$ is nothing more than the ($i$ times the) action of a probe scalar field on the double cone with boundary conditions corresponding to boundary couplings $g_1 = g+\frac{\delta g}{2}$ and $g_2 = g - \frac{\delta g}{2}$, evaluated to $o(\delta g^2)$. 

This exponential decay is precisely what we expected from~\eqref{E:parametricExpectation2} for the microcanonical form factor. All that remains is to fix the double cone moduli.

The cleanest way to proceed is to fix the energy directly and to adjust our boundary conditions slightly away from the double cone so that the two circles have $\beta_1 = \beta + i T$ and $\beta_2 = \beta - i T$ with $T\gg \beta$. Were it not for the constant energy constraint, we would be off-shell. With the constraint on total energy we find a constrained saddle where we are effectively dealing with the problem of a probe scalar field on top of the wormholes we reviewed in Section~\ref{sec:review}. For $T\gg \beta$ the scalar action leads to almost the same perturbation as the one we found above in~\eqref{E:decayOfDC} where $X^2$ is the value appropriate for the energy we have fixed. More generally, to $o(\delta g^2)$ we have a scalar action of the form $\beta_1 f(\beta_2/\beta_2) \delta g^2$, leading to a shift to the exponent of
\beq
\label{E:generalScalarS}
	-\pi X^2 T - 2 \beta \Delta E + O(\beta^2/T)\,,
\eeq
for some constant $\Delta E$ proportional to $\delta g^2$. This describes the exponential decay in real time along with a shift of the average energy due to the profile of the bulk scalar. In the spectrum of fluctuations around this wormhole there is a single light degree of freedom, a time translation zero mode of one boundary relative to the other, with a volume $\propto T$ and so we find a form factor
\beq
\label{E:holographicSFF}
	\left\langle \text{tr}\left( e^{-(\beta + i T) H(g+\delta g/2)}\right)\text{tr}\left( e^{-(\beta-i T)H(g-\delta g/2)}\right)\right\rangle_{\mathcal{C}} \sim T e^{ - \pi X^2 T-2\beta E + O(\delta g^3)}\,.
\eeq
This form factor is almost, but not quite the one appearing in~\eqref{E:parametricExpectation2}. There we considered energies $E_1 \approx E_2\approx E$, and did not have Boltzmann factors $\sim e^{-\beta H}$. Here we have Boltzmann factors as well as a constraint only on the total energy. However, the right-hand-side here comes from states with average energy $E$, leading to the Boltzmann factor $e^{-2\beta E}$. Had we introduced Boltzmann factors into the form factor appearing in~\eqref{E:parametricExpectation2}, then it would agree with the holographic result here~\eqref{E:holographicSFF}, with precisely the correct exponential decay and $T$-linear prefactor.

The match between the holographic form factor~\eqref{E:holographicSFF} and the expectation from the quantum chaos literature is the precision test in the title of this manuscript. 

Suppose that instead of fixing the energy directly that we pick $\beta_1$ and $\beta_2$ to be complex values such that the wormhole is a solution to Einstein's equations, and then perform an appropriate inverse Laplace transform. Doing so has the advantage that the wormhole so produced is a solution to the Einstein's equations rather than a constrained saddle. However there is a potential problem, that will become much more clear soon when we study specific examples where we can compute the scalar backreaction. Suppose $g$ is conjugate to a dimension $\Delta$ operator in the dual CFT. For double cones where boundary energy is much larger than its minimum possible value, by dimensional analysis we have
\beq
	X^2 =a E^{1+2(\Delta - d )} \delta g^2\,,
\eeq
for some constant $a$. To see the effect of the inverse Laplace transform with respect to $\beta=\text{Re}(\beta_1)=\text{Re}(\beta_2)$ with a Gaussian broadening of width $\Delta$, we need to go a bit off-shell by turning on a small amount of $\beta$. The total wormhole action is then, including the Einstein-Hilbert and scalar contributions to the action and the factors from the inverse Laplace transform,
\beq
	S_{\rm eff}=2\beta E' + \pi X^2 T - 2\beta E - \frac{\beta^2 \Delta^2}{2}\,.
\eeq
This is the effective action for a wormhole with energy $E'$ where we are fixing the energy to $E$ with a precision $\Delta$. With $\delta g = 0$, i.e.~$X^2 = 0$, extremizing with respect to the wormhole energy $E'$ leads to $\beta = 0$, but with $\delta g \neq 0$ the saddle point value is moved to
\beq
	\beta \approx - \frac{\pi a}{2}(1+2(\Delta -d))E^{2(\Delta-d)} \delta g^2 T\,,
\eeq
which is negative for a range of operator dimensions $\Delta > d-\frac{1}{2}$ including when the deformation is marginal. This is a potential problem, as the wormhole geometry near the first boundary is asymptotically
\beq
	ds^2 \approx e^{2\rho}\left(  (\beta+i T)^2 d\tau^2 + d\Omega_{d-1}^2\right) + d\rho^2\,,
\eeq
and similarly near the second boundary.
This is a complex metric that for negative $\beta$ violates the recently proposed Kontsevich-Segal criterion~\cite{Kontsevich:2021dmb,Witten:2021nzp} for an admissible complex metric, meaning the real part of the action of a matter field is not bounded below.  This pathology, which can be seen explicitly in our examples, is why we have taken the constrained saddle approach.

In the remainder of this Subsection we study examples where we can easily compute $X^2$ and the wormholes that compute the parametric form factor.

\subsubsection{Example: AdS$_3$ and detuning a marginal coupling}

We begin with a toy model, three-dimensional Einstein gravity coupled to a massless scalar field, 
\beq
	S = -\frac{1}{16\pi G}\int d^3x\sqrt{g}\left( R +2 - \frac{1}{2}(\partial\phi)^2\right) + S_{\rm bdy}\,,
\eeq
in units where the AdS radius is unity. In AdS$_3$ compactifications of string/M-theory like that arising for the D1/D5 system there are instead multiple moduli, with an effective action \\ $\frac{1}{32\pi G}\int d^3x \sqrt{g} G_{ab}(\varphi)\partial_{\mu} \varphi^a\partial^{\mu} \varphi^b$ where $a,b$ are indices running over the moduli of the compactification and $G_{ab}$ is the moduli space metric. We will study the toy model for now, and return to honest compactifications shortly. 

In radial gauge there are constrained saddle wormholes of fixed energy,
\beq
\label{E:AdS3wormhole}
	ds^2 = d\rho^2 + 4b^2\cosh^2(\rho)\left( \left( \frac{\beta_1 e^{\rho}+\beta_2 e^{-\rho}}{2\cosh(\rho)}\right)^2d\tau^2 + dx^2\right)\,,
\eeq
along with $\phi = g=\text{constant}$ and $\tau\sim \tau+1, x\sim x+1$. The boundary values of $\phi$ correspond to equal couplings $g$ for $\mathcal{O}$. The wormhole action is $\frac{b^2}{4\pi G}(\beta_1+\beta_2)$ so that the boundary energies are $E_1=E_2 = \frac{b^2}{4\pi G}$. 

Now we adjust the boundary conditions on $\phi$ so that $\lim_{\rho\to\pm \infty} \phi = g \pm \frac{\delta g}{2}$ with $\delta g$ small. Because $\delta g$ is small we may work in a probe approximation. We do so by introducing a formal expansion parameter $\varepsilon$, rescale $\delta g$ as $\delta g \to \varepsilon \delta g$, and expand the scalar and metric in powers of $\varepsilon$,
\begin{align}
\begin{split}
	\phi &= \varepsilon \phi^{(1)} + \varepsilon^3 \phi^{(3)} + O(\varepsilon^5)\,, 
	\\
	g_{ij} & = g_{ij}^{(0)} + \varepsilon^2 g_{ij}^{(2)} + O(\varepsilon^4)\,.
\end{split}
\end{align}
Really $\varepsilon$ is a standby for the small difference in boundary couplings $\delta g$. Expanding the scalar and Einstein's equations in powers of $\varepsilon$, to leading order in $\varepsilon$ in the scalar and Einstein's equations we have
\begin{align}
\begin{split}
	D_{\mu}^{(0)}D^{\mu\,(0)}\phi^{(1)} & = 0\,,
	\\
	R_{ij}^{(2)} -\left(  \frac{R^{(2)}}{2}g^{(0)}_{ij} +\frac{R^{(0)}}{2}g^{(2)}_{ij}\right)-g^{(2)}_{ij} & = T_{ij}^{(2)}\,,
\end{split}
\end{align}
where $D_{\mu}^{(0)}$ refers to the covariant derivative constructed from the background metric $g^{(0)}$, $R^{(0)}$ is the scalar curvature of the background, $R_{ij}^{(2)}$ and $R^{(2)}$ refer to the $O(\varepsilon^2)$ terms in the Ricci and scalar curvatures, and 
\beq
	T_{ij}^{(2)} =\frac{1}{2} \partial_i \phi^{(1)}\partial_j \phi^{(2)} - \frac{1}{4} (\partial_k\phi^{(1)} \partial_l \phi^{(1)} g^{kl\,(0)})g_{ij}^{(0)}
\eeq
is the $O(\varepsilon^2)$ term in the stress tensor.

We make an ansatz for the solution that respects the symmetries of the perturbed boundary conditions, namely translation invariance in the boundary directions $(\tau,x)$ and parity,
\beq
	\phi^{(1)} = f(\rho)\,, \qquad g_{ij}^{(2)}dx^i dx^j = h_{\tau\tau}(\rho)d\tau^2 + h_{xx}(\rho)dx^2\,.
\eeq
The scalar equation of motion is simply a conservation equation, with
\beq
	f'(\rho) = \frac{q}{\sqrt{g}}\,,
\eeq
for some constant $q$. Integrating over $\rho$ and enforcing the boundary conditions fixes
\beq
	q = \frac{2b^2(\beta_1-\beta_2)}{\ln \left( \frac{\beta_1}{\beta_2}\right)}\,\delta g\,,
\eeq
and gives
\beq
	\phi^{(1)} = \frac{\ln \left(\frac{1}{\beta_1\beta_2}\left( \frac{\beta_1 e^{\rho}+\beta_2 e^{-\rho}}{2\cosh(\rho)}\right)^2\right)}{2\ln \left( \frac{\beta_1}{\beta_2}\right)}\,\delta g\,.
\eeq
The $O(\varepsilon^2)$ term in the action is just computed by the scalar action, evaluated on the solution for $\phi^{(1)}$.\footnote{The $O(\varepsilon^2)$ term in the variation of the gravitational part of the action vanishes, by the usual observation in the probe limit that the background we are expanding around is stationary against perturbations that do not change the asymptotic metrics.} It is
\beq
	S^{(2)} = \frac{q\delta g}{32\pi G} =  \frac{b^2}{16\pi G}\frac{\beta_1-\beta_2}{\ln \left( \frac{\beta_1}{\beta_2}\right)}\delta g^2\,.
\eeq

The boundary energies are no longer equal on account of the scalar profile. Whether by computing the boundary stress tensors or by evaluating $\frac{\partial S}{\partial\beta_1}$ and $\frac{\partial S}{\partial \beta_2}$, we have
\begin{align}
\begin{split}
	E_1 & = \frac{b^2}{4\pi G}\left( 1 +\frac{\delta g^2}{4}\left( \frac{1}{\ln^2\left( \frac{\beta_1}{\beta_2}\right)} + \frac{\beta_2-\beta_1}{\beta_1 \ln \left( \frac{\beta_1}{\beta_2}\right)}\right) \right)+O(\delta g^4)\,,
		\\
	E_2 & = \frac{b^2}{4\pi G}\left( 1 -\frac{\delta g^2}{4}\left( \frac{1}{\ln^2\left( \frac{\beta_1}{\beta_2}\right)} + \frac{\beta_1-\beta_2}{\beta_2 \ln \left( \frac{\beta_1}{\beta_2}\right)}\right) \right)+O(\delta g^4)\,.
\end{split}
\end{align}
As a result we can stabilize the wormhole by fixing the total energy, but for general $\beta_1$ and $\beta_2$ this determines the wormhole modulus $b$ in a way that depends on $\beta_1, \beta_2$, and $\delta g$. 

Let us now take $\beta_1=\beta+i T$ and $\beta_2 = \beta - i T$ with $T\gg \beta$. The scalar action becomes
\beq
	S^{(2)} = \frac{b^2\delta g^2}{8\pi^2 G} \,T + \frac{b^2\delta g^2}{4\pi^2G}\,\beta + O\left( \frac{\beta^2}{T}\right)\,,
\eeq
consistent with our expectation in~\eqref{E:generalScalarS}. The term linear in $T$ is $\pi X^2 T$ where $X^2$ is the relevant response function for the BTZ black hole with energy $E = \frac{b^2}{4\pi G}$, while the term linear in $\beta$ is a shift in the energy. The average boundary energy in this limit is
\beq
	\frac{E_1 + E_2}{2} = \frac{b^2}{4\pi G}\left( 1 + \frac{\delta g^2}{8\pi^2}\right)+O\!\left( \frac{\beta}{T}, \delta g^4\right)\,,
\eeq
so that fixing the average energy to be $E>0$ fixes the parameter $b$. The stabilized wormhole action then reads
\beq
	S = 2\beta E + \frac{E\delta g^2}{2\pi} T + O\!\left( \frac{\beta^2}{T},\delta g^4\right)\,.
\eeq
so that after including the twist zero mode the dual form factor is approximately
\beq
	\left\langle \text{tr}\left( e^{-(\beta+i T)H(g+\delta g/2)}\right)\text{tr}\left( e^{-(\beta-iT)H(g-\delta g/2)}\right)\right\rangle_{\mathcal{C}} \sim T e^{-\frac{E\delta g^2}{2\pi} T - 2\beta E}\,.
\eeq

At the beginning of this Subsection we gave a very general argument that the late time form factor obtained from the wormhole decays at the rate $\pi X^2$ expected from the quantum chaos literature. This instance gives an explicit example of that argument in action. The decay rate $\frac{E \,\delta g^2}{2\pi}$ comes from the scalar action as $\beta \to 0$, in which case the wormhole becomes the double cone and the rate can be reinterpreted as a cross-horizon zero-frequency function of the operator conjugate to $\phi$. But, as we discussed above, that implies that the rate is nothing more than the quantity $\pi X^2$ in~\eqref{E:simpleX2}, namely the rate predicted by~\eqref{E:parametricExpectation2}. 

Now let us consider a string compactification with multiple moduli, like the D1/D5 system or MSW strings~\cite{Maldacena:1997de}. The corresponding computation is almost the same as in the toy model. Let us expand around the point in moduli space $\varphi^a = g^a$ for some vector of marginal couplings $g^a$, and diagonalize the moduli space metric at that point. Then in the probe limit the moduli decouple, giving a number of independent massless scalar fields. To quadratic order in the difference in couplings $\delta g^a$ we then have
\beq
	S^{(2)} = \frac{b^2}{16\pi G}  \frac{\beta_1-\beta_2}{\ln \left( \frac{\beta_1}{\beta_2}\right)}\,G_{ab}(g) \delta g^a \delta g^b\,,
\eeq
leading to the correct decay in the real-time form factor when we turn on perturbations in the couplings for spacelike directions of the moduli space. Effectively we have $\delta g^2 \to G_{ab}\delta g^a\delta g^b$.

One can repeat this computation in a spacetime dimension other than three, with nearly identical results for the scalar action. In three dimensions however one can find simple expressions for the backreaction of the metric. We find for $\beta_1,\beta_2>0$
\begin{align}
\nonumber
	h_{\tau\tau} & = \frac{b^2 \delta g^2}{2\ln\left( \frac{\beta_1}{\beta_2}\right)^2} 	\left( (\beta_1 e^{\rho}+\beta_2 e^{-\rho})^2\ln \left( \frac{2\cosh\rho}{\beta_1 e^{\rho}+\beta_2 e^{-\rho}}\right)+ (\beta_1 e^{\rho}+\beta_2 e^{-\rho})(\beta_1 e^{\rho} \ln\beta_1+ \beta_2 e^{-\rho}\ln \beta_2) \right)\,,
\\
	h_{xx} & = -\frac{b^2\delta g^2}{\ln \left( \frac{\beta_1}{\beta_2}\right)^2}\cosh\rho\left( 2\cosh\rho\ln \left(\frac{2\cosh\rho}{\beta_1e^{\rho}+\beta_2e^{-\rho}}\right) +e^{\rho}\ln\beta_1+e^{-\rho}\ln\beta_2 \right)\,.
\end{align}

Above, we discussed how it is convenient to fix the energy directly, leading to a constrained saddle wormhole. Suppose that instead we pick $\beta_1$ and $\beta_2$ to be complex values tuned so as to put the wormhole on-shell. Without the scalar profile that leads to $\beta_1=-\beta_2$, which for $\text{Re}(\beta_1),\text{Re}(\beta_2)\geq 0$ implies $\beta_1 = iT, \beta_2=-iT$. With the scalar profile, we find that this background becomes a solution to the $\rho\rho$ component of Einstein's equations when
\beq
	b^2\left( \beta_1+\beta_2+ \frac{\beta_1-\beta_2}{4\ln \left( \frac{\beta_1}{\beta_2}\right)} \delta g^2 + O(\delta g^4)\right) = 0\,,
\eeq
which is equivalent to $\frac{\partial S_{\rm wormhole}}{\partial b} = 0$. The solution is
\beq
	\beta_1=  -\frac{\delta g^2}{4\pi }T + i T + O(\delta g^4)\,, \qquad \beta_2 = -\frac{\delta g^2}{4\pi}T - i T +O(\delta g^4)\,.
\eeq
Note that the scalar profile pushes the saddle point value of $\beta$, the real part of $\beta_1$ and $\beta_2$, to a small negative value. For this reason we find it preferable to instead fix the energy directly, and so we work with a constrained saddle.

\subsubsection{Example: AdS$_5\times \mathbb{S}^5$ wormholes and detuning the Yang-Mills coupling}

Now let us consider a richer example relevant for the paradigmatic example of the AdS/CFT correspondence, the duality between $\mathcal{N}=4$ super-Yang Mills with gauge group $SU(N)$ and type IIB string theory on AdS$_5\times \mathbb{S}^5$ with $N$ units of five-form flux. We wish to consider wormholes connecting two asymptotic regions with $\mathbb{S}^1\times\mathbb{S}^3$ boundary, and slightly detuned values of the Yang-Mills coupling. The 5d effective action is
\beq
	S = -\frac{1}{16\pi G_5}\int d^5x \sqrt{g} \left( R + 12 - \frac{1}{2}(\partial\Phi)^2\right)\,,
\eeq 
where we have normalized the AdS$_5$ radius to unity, $G_5 = \frac{\pi}{2N^2}$, and $\Phi$ is the 5d dilaton, dual to the $\mathcal{N}=4$ Yang-Mills coupling. 

The wormhole we will perturb around is
\beq
	ds^2 = d\rho^2 + \frac{b^2\cosh(2\rho)-1}{2}\left( \left( \frac{\beta_1e^{2\rho}+\beta_2 e^{-2\rho}}{2\left( \cosh(2\rho)-\frac{1}{b^2}\right)}\right)^2d\tau^2 + d\Omega_3^2\right)\,,
\eeq
along with constant $\Phi = g$. The unperturbed wormhole action is $(\beta_1+\beta_2)E$ and the boundary energies are $E=E_1 = E_2 = \frac{3\pi b^4}{32G}$ (in a scheme where $E_0 = \frac{3\pi}{32G}$ is the vacuum energy, and also the energy of the lightest black hole) so that fixing the energy fixes $b$, and the geometry is regular only for $b>1$. As in the AdS$_3$ example we study probe fluctuations on top of this background, with
\begin{align}
\begin{split}
	\Phi & = \varepsilon \Phi^{(1)} + O(\varepsilon^3)\,,
	\\
	g_{ij} & = g_{ij}^{(0)} + \varepsilon^2 g_{ij}^{(2)} + O(\varepsilon^4)\,.
\end{split}
\end{align}
The radial derivative of the scalar perturbation is
\beq
	\partial_{\rho}\Phi^{(1)} = q \frac{\sqrt{g_{\mathbb{S}^3}}}{\sqrt{g}} = \frac{8q}{b^2(b^2\cosh(2\rho)-1)(\beta_1e^{2\rho} + \beta_2 e^{-2\rho})}\,,
\eeq
for some constant $q$ fixed by the boundary conditions $\lim_{\rho\to \pm\infty} \Phi^{(1)} = \pm \frac{\delta g}{2}$, but the precise expressions for $\Phi^{(1)}$ and the metric perturbations are quite complicated, and we omit them here. The result for the perturbation of the action however is relatively simple, and reads
\begin{align}
	S^{(2)} &=\frac{1}{32\pi G_5}\int d^5x \sqrt{g} (\partial\Phi)^2 =  \frac{q \,\delta g}{32\pi G_5} \text{vol}(\mathbb{S}^3)
	\\
	\nonumber
	& =  \frac{\pi}{64G_5} \frac{b^2\sqrt{b^4-1}(b^4(\beta_1-\beta_2)^2+4\beta_1\beta_2)}{b^2(\sqrt{b^4-1}(\beta_1-\beta_2)\ln\left(\frac{\beta_1}{\beta_2}\right)+(\beta_1+\beta_2)(2\,\text{arccsc}(b^2)+\pi ))-2\pi \sqrt{\beta_1\beta_2(b^4-1)}}\,\delta g^2\,.
\end{align}
Taking $\beta_1 = \beta+iT$ and $\beta_2 = \beta-i T$ and $T\gg \beta$ we then have
\beq
	S = 2\beta E + \frac{b^2(b^2-1)\delta g^2}{64G_5}\,T + O(\delta g^4)\,,
\eeq
where the average energy is
\beq
	E = \frac{3\pi b^4}{32G_5}+\frac{b^4\sqrt{b^4-1}(2\sqrt{b^4-1}+2\,\text{arccsc}(b^2)+\pi)}{128\pi (b^2+1)^2G_5} \,\delta g^2+ O(\delta g^4)\,.
\eeq
This wormhole is stabilized by fixing this average energy so that $b^4 = \frac{E}{E_0} + O(\delta g^2)$, and accordingly the form factor is approximately
\beq
	\left\langle \text{tr}\left( e^{-(\beta+iT)H(g+\delta g/2)}\right)\text{tr}\left( e^{-(\beta-iT)H(g-\delta g/2)}\right)\right\rangle_{\mathcal{C}} \sim T e^{- 2\beta E - \pi X^2 T}
\eeq
where the decay rate is
\beq
	\pi X^2 = \frac{\pi N^2\sqrt{\frac{E}{E_0}}\left(\sqrt{\frac{E}{E_0}}-1\right)}{32} \,\delta g^2 \,.
\eeq
This rate is precisely the one in~\eqref{E:simpleX2} expected from the chaos literature: the computation of the rate from the wormhole is the same one that leads to the zero-frequency cross-horizon two-point function appearing in~\eqref{E:simpleX2}.

As with the AdS$_3$ wormholes discussed above, we could make these wormholes solutions to the Einstein's equations by choosing $\beta_1$ and $\beta_2$ appropriately. The solution that does the job is
\beq
	\beta_1 = i T - \frac{N^2\left( 2-\sqrt{\frac{E_0}{E}}\right)}{128\pi E_0}\delta g^2 T + O(\delta g^4)\,, \qquad \beta_2 = i T - \frac{N^2\left( 2 - \sqrt{\frac{E_0}{E}}\right)}{128\pi E_0}\delta g^2T + O(\delta g^4)\,,
\eeq
which has negative $\text{Re}(\beta_1)$ and $\text{Re}(\beta_2)$. This wormhole solves the Einstein's equations but violates the Kontsevich-Segal criterion as expected.  Again, this is why we have opted for constraining the energy instead.

\section{Correlations between theories at different $N$} 
\label{sec:Ncorr}

\subsection{Wormholes and branes}

Next we turn to a more drastic comparison between two copies of holographic CFTs, namely where we look at correlations between replicas with a different number of branes. For example we will consider two replicas of $\mathcal{N}=4$ super-Yang Mills theory, with gauge group $SU(N)$ versus $SU(N+1)$, although our analysis will generalize to one comparing theories with gauge groups $SU(N+k)$ for $k \sim O(1)$. The role of correlations between holographic theories of different gauge groups was proposed as a question of interest in the recent work of Schlenker and Witten~\cite{Schlenker:2022dyo}; here we can study the imprint of those correlations on the spectral form factor. 

Since the $\mathcal{N} = 4$ SYM action goes as $S \sim N^2$, modifying $N$ changes both the leading profile of the density of states and so the mean level spacing which goes like $\sim e^{-N^2}$. This means that the standard results of parametric correlations in chaotic systems~\cite{Simons:1993zza, simons1993universalities, guhr1998random} that we used in the last Section do not apply, so there is no presently known quantitative prediction for a form factor constructed from $\langle Z_N Z_{N+1}\rangle$.  We will comment on this more below.  Nonetheless, we can proceed holographically, and see what we find.

To begin we must review how to construct the wormholes that contribute to the spectral form factor of theories with a string theory dual. For concreteness, let us work in the standard setting of Euclidean AdS$_5 \times \mathbb{S}^5$ with $\mathbb{S}^1 \times \mathbb{S}^3$ boundary, although our analysis will generalize naturally to other settings.  We obtained wormholes of this sort in our previous work~\cite{Cotler:2021cqa}, which can be rendered bona fide saddles in supergravity by introducing an energy constraint as in Section~\ref{sec:fixing}. The 10d Einstein frame metric, dilaton, and RR four-form potentials specifying the wormholes are
\begin{align}
ds_E^2 &= \frac{b^2 \, \cosh(2\rho) - 1}{2}\left(\left(\frac{\beta_1 e^{2\rho} + \beta_2 e^{-2\rho}}{2(\cosh(2\rho) - \frac{1}{b^2})}\right)^2 d\tau^2 + d\Omega_3^2\right) + d\rho^2 + d\Omega_5^2\,, \quad e^{\Phi} = g_s\,, \nonumber \\
C_4 &= - i \left[\frac{b^4}{16}(\beta_1 e^{4\rho} - \beta_2 e^{-4\rho} + 4 (\beta_1 + \beta_2) \rho) - \frac{b^2}{4}(\beta_1 e^{2\rho} - \beta_2 e^{-2\rho})\right] d\tau \wedge \text{vol}_{\mathbb{S}^3} + (\text{angular})\,, \nonumber
\end{align}
where $b > 1$ and $\partial_\rho C_{\tau a b c} = - 4 i \, \varepsilon_{abc}$ with $a,b,c$ being angles on the 3-sphere. Fixing the total boundary energy fixes $b$. We are working in units where the AdS radius is unity, and the string length is $\frac{1}{l_s^4} = 4\pi N$. 

This background connects two Euclidean AdS$_5\times\mathbb{S}^5$ throats, each with $N$ units of RR five-form flux. To study a form factor where the two replicas have different values of $N$ we must introduce a source for RR flux, i.e.~a D3 brane in the wormhole, which will lead to an asymmetry in the asymptotic RR fluxes. In our previous work we studied various 3-brane instantons in these wormholes, mostly focusing on D3-$\overline{\text{D3}}$ pairs from the point of view of studying brane nucleation. There we found (see also~\cite{Mahajan2021wormholes}) that, when taking the continuation required to study the real-time form factor, that these wormholes are always stable against obvious nucleation channels at sufficiently large time $T$.

Now let us insert a single D3 brane, rather than a brane pair. In particular consider a D3 brane extended along $\mathbb{S}^1 \times \mathbb{S}^3$ but at a particular point on the $\mathbb{S}^5$, and at a particular value of $\rho$ to be determined.  We denote the location along $\rho$ by $\rho_0$. Crucially there is a saddle point value for $\rho_0$. The action of our desired D3 brane in the probe regime is~\cite{Cotler:2021cqa}
\begin{align}
\begin{split}
	S_{D3} &= \mu_3 \left(\int d^4 \sigma \sqrt{P[G_E]} -i \int P[C_4]\right)
	\\
	&= \frac{b^4 N \,\text{Vol}(\mathbb{S}^3)}{32\pi^2} \left(2 \beta_2 e^{-4 \rho_0} - (\beta_1 + \beta_2)(4\rho_0 - 1) + \frac{2}{b^2}(\beta_1 e^{2\rho_0} - 3 \beta_2 e^{-2\rho_0})\right)\,,
\end{split}
\end{align}
where $\int d^4 \sigma \sqrt{P[G_E]}$ is the DBI term and $-i \int P[C_4]$ is the WZW term.  Examining the second line and thinking of it as an effective potential $V_{\text{eff}}(\rho_0)$, one can check that there is a particular value of $\rho_0$ where the gravitational attraction and RR repulsion terms balance. The exact expression for the minimizing $\rho_0$ which we will call $\rho_0(b, \beta_1, \beta_2)$ is not particularly illuminating.  Nonetheless, we are interested in $\beta_1 \to \beta + i T$ and $\beta_2 \to \beta - i T$ for $T \gg \beta \gg 1$ in which case
\begin{equation}
\rho_0(b, \beta + iT, \beta - iT) = \frac{1}{2} \,\log
   \left(\frac{(-1)^{1/3}-\left(-b^2-\sqrt{b^4-1}\right)^{2/3}}{(b^2+\sqrt{b^4-
   1})^{1/3}}\right) + O(T^{-1})\,.
\end{equation} 
Plugging this into the D3 brane action and considering large $b$ to simplify the expression, we find
\begin{align}
\label{E:D3final1}
S_{D3}(b, \beta + iT, \beta - i T) &= N T \cdot \frac{3}{8}\left(\frac{i \,b^8}{4}\right)^{1/3} \left(1 + O\left(b^{-\frac{4}{3}}\right)\right)\,.
\end{align}
If we had $N\gg k$ D3 branes, we would find $k$ times this result.  

Combining the above with our previous results, for $T \gg \beta \gg 1$ and exchanging $b$ for an average energy $E$ we find
\begin{equation}
\label{E:Ncompare1}
	\left\langle\left( \text{tr}\big(e^{- (\beta + i T) H_{N+1}}\right)\,\text{tr}\left(e^{- (\beta - i T) H_N}\right)\right\rangle_{\mathcal{C}} \sim T \, e^{- 2 \beta \, E -  N T \, \frac{3}{2}\,i^{1/3}\, \left(\frac{E}{N^2}\right)^{2/3}\left(1 + O\left(\left(\frac{E}{N^2}\right)^{-\frac{1}{3}}\right)\right)}\,,
\end{equation}
where $E =O(N^2)$ lies above the black hole threshold. We see that the ramp decays exponentially in $T$ with a decay time $\sim 1/N$ with a wildly oscillating phase. So correlations between two copies of $\mathcal{N} = 4$ SYM with gauge groups $SU(N)$ versus $SU(N+1)$ are exponentially damped with $N$, so that inter-replica long-range level repulsion is parametrically small.

Similar analyses may be performed for the standard stringy examples of AdS/CFT, including ABJM  and the $\mathcal{N}=(2,0)$ theory. In each case one finds the same qualitative physics, namely an exponentially decaying, wildly oscillating form factor.

\subsection{Comments on random matrix theory correlations at different $N$}

Existing results for parametric correlations in chaotic systems~\cite{Simons:1993zza, simons1993universalities, guhr1998random} are insufficient to corroborate our result~\eqref{E:Ncompare1}.  In particular, we have deviated from the literature on parametric correlations in chaotic systems and random matrix theory by studying replicas with different values of the leading density of states. However, as we stressed in the Introduction, studying correlations between holographic CFTs with different values of $N$ is not so drastic as one might think. Zooming out from the near-horizon limit, black holes in asymptotically flat string/M-theory are characterized by superselection sectors with fixed flux, and from that point of view, our~\eqref{E:Ncompare1} is examining tiny correlations between different superselection sectors of fixed RR five-form flux at energies close to the black hole threshold (so that there are large Euclidean AdS throats).

Ordinarily, chaotic quantum systems with conserved charges are approximated in random matrix theory as having independent spectral statistics in each charge sector.  That is, each sector is associated with an independent random matrix distribution.  This approximation does not fully reflect the reality of chaotic quantum systems, which have small but nonzero correlations between charge sectors.  At present, there is not a clean theory of such tiny correlations.  One plausible route would be to study universal features of multi-matrix models with mulitrace deformations, where each matrix corresponds to a symmetry sector and the mulitrace deformations (weakly) couple the sectors.  Another approach would be to generalize the sigma model analysis of Altshuler and Simons~\cite{Simons:1993zza, simons1993universalities}, by giving certain fields in their action a VEV.

We note that in our previous work on pure AdS$_3$ gravity~\cite{cotler2020ads}, the spectral form factor for BTZ black holes contained non-trivial correlations between sectors of different angular momentum.  While this feature was present in our results, we did not explore it further.  To study these correlations, one would consider our full result for the spectral form factor and take the large angular momentum limit, followed by the limit of large Lorentzian time $T$.  This would plausibly produce a digestible formula for small correlations between superselection sectors of a chaotic system.

We conclude by making some more concrete remarks about the basic structure of the answer~\eqref{E:Ncompare1}. First, it is easy to see why~\eqref{E:Ncompare1} must be complex.  The usual spectral form factor
\begin{equation}
\big\langle \text{tr}\big(e^{-(\beta + i T) H(N)}\big)\,\text{tr}\big(e^{-(\beta - i T) H(N)}\big) \big\rangle_{\mathcal{C}}
\end{equation}
is manifestly real since it equals its own complex conjugate. By contrast,
\begin{equation}
\big\langle \text{tr}\big(e^{-(\beta + i T) H(N)}\big)\,\text{tr}\big(e^{-(\beta + i T) H(N+1)}\big) \big\rangle_{\mathcal{C}}
\end{equation}
is not equal to its complex conjugate, and so is complex.  The level spacings of $H(N)$ go as $\sim e^{-N^2}$ whereas for $H(N+1)$ they go as $\sim e^{-(N+1)^2}$, and so there are vastly more energy levels in a fixed energy window in the latter case.  As such, there is no way that eigenvalues from $H(N)$ and $H(N+1)$ can pair up in the spectral form factor to cancel complex phases.  This is in contrast to what happened in our previous analyses of the spectral form factor with slightly different $g_{\text{YM}}$ couplings between the two copies.  There the mean level spacing stayed fixed and the eigenvalues (on average) did not move around due to the perturbation; this enabled cancellations of complex phases in our answer.

In the $N$ vs.~$N+1$ setting, a simple picture for the form factor
$$\Big\langle \text{tr}\!\left( e^{-(\beta + i T)H(N)}\right)\text{tr}\!\left(e^{-(\beta-iT)H(N+1)}\right)\Big\rangle_{\mathcal{C}} \sim e^{-\gamma \frac{\partial S}{\partial N} T - 2 \beta E}$$
is to consider an $O(1)$-sized window of energies above the black hole threshold.  The spectrum of the replica with gauge group $SU(N+1)$ has $\sim e^{(N+1)^2}$ eigenvalues in this window with a level spacing of $e^{-(N+1)^2}$, while there are only $e^{N^2}$ eigenvalues of the $SU(N)$ theory in the same range.  If $\lambda$ and $\lambda'$ are adjacent eigenvalues in the $SU(N)$ spectrum, then on average they will be separated by $\sim e^{\frac{\partial S}{\partial N}}$ eigenvalues of the $SU(N+1)$ spectrum, where $\frac{\partial S}{\partial N}$ is the derivative of the coarse-grained entropy. Since $N$ can be regarded as a charge in the 10d string theory description, $\frac{\partial S}{\partial N}$ may be regarded as a chemical potential.  The influence of the $\sim e^{\frac{\partial S}{\partial N}}$ interstitial eigenvalues apparently conspires to decorrelate the embedded $SU(N)$ eigenvalues by $\sim e^{-\frac{\partial S}{\partial N}}$, as evidenced by the $N$-dependence of~\eqref{E:Ncompare1}.

\subsection*{Acknowledgments}
We thank Alex Altland, Tom Hartman, Nick Hunter-Jones, Juan Maldacena, Don Marolf, Moshe Rozali, Jorge Santos, Steve Shenker, Douglas Stanford, and Edward Witten for valuable discussions.  JC is supported by a Junior Fellowship from the Harvard Society of Fellows, the Black Hole Initiative, as well as in part by the Department of Energy under grant {DE}-{SC0007870}.  KJ is supported in part by NSERC.

\bibliography{refs}

\providecommand{\href}[2]{#2}\begingroup\raggedright\begin{thebibliography}{10}

\bibitem{witten1999connectedness}
E.~Witten and S.-T. Yau, {\it {Connectedness of the boundary in the AdS/CFT
  correspondence}},  \href{http://xxx.lanl.gov/abs/hep-th/9910245}{{\tt
  hep-th/9910245}}.

\bibitem{maldacena2004wormholes}
J.~Maldacena and L.~Maoz, {\it {Wormholes in AdS}},  {\em Journal of High
  Energy Physics} {\bf 2004} (2004), no.~02 053,
  [\href{http://xxx.lanl.gov/abs/hep-th/0401024}{{\tt hep-th/0401024}}].

\bibitem{Jensen:2016pah}
K.~Jensen, {\it {Chaos in AdS$_2$ Holography}},  {\em Phys. Rev. Lett.} {\bf
  117} (2016), no.~11 111601, [\href{http://xxx.lanl.gov/abs/1605.06098}{{\tt
  1605.06098}}].

\bibitem{Engelsoy:2016xyb}
J.~Engelsoy, T.~G. Mertens, and H.~Verlinde, {\it {An investigation of
  AdS$_{2}$ backreaction and holography}},  {\em JHEP} {\bf 07} (2016) 139,
  [\href{http://xxx.lanl.gov/abs/1606.03438}{{\tt 1606.03438}}].

\bibitem{Maldacena:2016upp}
J.~Maldacena, D.~Stanford, and Z.~Yang, {\it {Conformal symmetry and its
  breaking in two dimensional Nearly Anti-de-Sitter space}},  {\em PTEP} {\bf
  2016} (2016), no.~12 12C104, [\href{http://xxx.lanl.gov/abs/1606.01857}{{\tt
  1606.01857}}].

\bibitem{Stanford:2019vob}
D.~Stanford and E.~Witten, {\it {JT gravity and the ensembles of random matrix
  theory}},  {\em Adv. Theor. Math. Phys.} {\bf 24} (2020), no.~6 1475--1680,
  [\href{http://xxx.lanl.gov/abs/1907.03363}{{\tt 1907.03363}}].

\bibitem{maldacena2019two}
J.~Maldacena, G.~J. Turiaci, and Z.~Yang, {\it {Two dimensional Nearly de
  Sitter gravity}},  \href{http://xxx.lanl.gov/abs/1904.01911}{{\tt
  1904.01911}}.

\bibitem{cotler2019low}
J.~Cotler, K.~Jensen, and A.~Maloney, {\it {Low-dimensional de Sitter quantum
  gravity}},  \href{http://xxx.lanl.gov/abs/1905.03780}{{\tt 1905.03780}}.

\bibitem{Maxfield:2020ale}
H.~Maxfield and G.~J. Turiaci, {\it {The path integral of 3D gravity near
  extremality; or, JT gravity with defects as a matrix integral}},  {\em JHEP}
  {\bf 01} (2021) 118, [\href{http://xxx.lanl.gov/abs/2006.11317}{{\tt
  2006.11317}}].

\bibitem{Witten:2020wvy}
E.~Witten, {\it {Matrix Models and Deformations of JT Gravity}},  {\em Proc.
  Roy. Soc. Lond. A} {\bf 476} (2020), no.~2244 20200582,
  [\href{http://xxx.lanl.gov/abs/2006.13414}{{\tt 2006.13414}}].

\bibitem{Giombi:2008vd}
S.~Giombi, A.~Maloney, and X.~Yin, {\it {One-loop Partition Functions of 3D
  Gravity}},  {\em JHEP} {\bf 08} (2008) 007,
  [\href{http://xxx.lanl.gov/abs/0804.1773}{{\tt 0804.1773}}].

\bibitem{cotler2020ads}
J.~Cotler and K.~Jensen, {\it {AdS$_3 $ gravity and random CFT}},  {\em Journal
  of High Energy Physics} {\bf 2021} (2021), no.~4
  [\href{http://xxx.lanl.gov/abs/2006.08648}{{\tt 2006.08648}}].

\bibitem{Chandra:2022bqq}
J.~Chandra, S.~Collier, T.~Hartman, and A.~Maloney, {\it {Semiclassical 3D
  gravity as an average of large-c CFTs}},
  \href{http://xxx.lanl.gov/abs/2203.06511}{{\tt 2203.06511}}.

\bibitem{Eberhardt:2022wlc}
L.~Eberhardt, {\it {Off-shell Partition Functions in 3d Gravity}},
  \href{http://xxx.lanl.gov/abs/2204.09789}{{\tt 2204.09789}}.

\bibitem{Belin:2020hea}
A.~Belin and J.~de~Boer, {\it {Random statistics of OPE coefficients and
  Euclidean wormholes}},  {\em Class. Quant. Grav.} {\bf 38} (2021), no.~16
  164001, [\href{http://xxx.lanl.gov/abs/2006.05499}{{\tt 2006.05499}}].

\bibitem{Belin:2021ryy}
A.~Belin, J.~de~Boer, and D.~Liska, {\it {Non-Gaussianities in the Statistical
  Distribution of Heavy OPE Coefficients and Wormholes}},
  \href{http://xxx.lanl.gov/abs/2110.14649}{{\tt 2110.14649}}.

\bibitem{Anous:2021caj}
T.~Anous, A.~Belin, J.~de~Boer, and D.~Liska, {\it {OPE statistics from
  higher-point crossing}},  \href{http://xxx.lanl.gov/abs/2112.09143}{{\tt
  2112.09143}}.

\bibitem{Saad:2018bqo}
P.~Saad, S.~H. Shenker, and D.~Stanford, {\it {A semiclassical ramp in SYK and
  in gravity}},  \href{http://xxx.lanl.gov/abs/1806.06840}{{\tt 1806.06840}}.

\bibitem{Saad:2019lba}
P.~Saad, S.~H. Shenker, and D.~Stanford, {\it {JT gravity as a matrix
  integral}},  \href{http://xxx.lanl.gov/abs/1903.11115}{{\tt 1903.11115}}.

\bibitem{cotler2020ads2}
J.~Cotler and K.~Jensen, {\it {AdS$_3$ wormholes from a modular bootstrap}},
  {\em Journal of High Energy Physics} {\bf 2020} (2020), no.~11
  [\href{http://xxx.lanl.gov/abs/2007.15653}{{\tt 2007.15653}}].

\bibitem{affleck1981constrained}
I.~Affleck, {\it On constrained instantons},  {\em Nuclear Physics B} {\bf 191}
  (1981), no.~2 429--444.

\bibitem{Schlenker:2022dyo}
J.-M. Schlenker and E.~Witten, {\it {No Ensemble Averaging Below the Black Hole
  Threshold}},  \href{http://xxx.lanl.gov/abs/2202.01372}{{\tt 2202.01372}}.

\bibitem{eberhardt2021summing}
L.~Eberhardt, {\it {Summing over Geometries in String Theory}},
  \href{http://xxx.lanl.gov/abs/2102.12355}{{\tt 2102.12355}}.

\bibitem{guhr1998random}
T.~Guhr, A.~M{\"u}ller-Groeling, and H.~A. Weidenm{\"u}ller, {\it Random-matrix
  theories in quantum physics: common concepts},  {\em Physics Reports} {\bf
  299} (1998), no.~4-6 189--425,
  [\href{http://xxx.lanl.gov/abs/cond-mat/9707301}{{\tt cond-mat/9707301}}].

\bibitem{Simons:1993zza}
B.~D. Simons and B.~L. Altshuler, {\it {Universal velocity correlations in
  disordered and chaotic systems}},  {\em Physical Review Letters} {\bf 70}
  (1993) 4063--4066.

\bibitem{simons1993universalities}
B.~Simons and B.~Altshuler, {\it Universalities in the spectra of disordered
  and chaotic systems},  {\em Physical Review B} {\bf 48} (1993), no.~8 5422.

\bibitem{Cotler:2021cqa}
J.~Cotler and K.~Jensen, {\it {Wormholes and black hole microstates in
  AdS/CFT}},  {\em JHEP} {\bf 09} (2021) 001,
  [\href{http://xxx.lanl.gov/abs/2104.00601}{{\tt 2104.00601}}].

\bibitem{cotler2020gravitational}
J.~Cotler and K.~Jensen, {\it {Gravitational Constrained Instantons}},  {\em
  Phys. Rev. D} {\bf 104} (2021) 081501,
  [\href{http://xxx.lanl.gov/abs/2010.02241}{{\tt 2010.02241}}].

\bibitem{Kapec:2019ecr}
D.~Kapec, R.~Mahajan, and D.~Stanford, {\it {Matrix ensembles with global
  symmetries and \textquoteright{}t Hooft anomalies from 2d gauge theory}},
  {\em JHEP} {\bf 04} (2020) 186,
  [\href{http://xxx.lanl.gov/abs/1912.12285}{{\tt 1912.12285}}].

\bibitem{Kontsevich:2021dmb}
M.~Kontsevich and G.~Segal, {\it {Wick Rotation and the Positivity of Energy in
  Quantum Field Theory}},  {\em Quart. J. Math. Oxford Ser.} {\bf 72} (2021),
  no.~1-2 673--699, [\href{http://xxx.lanl.gov/abs/2105.10161}{{\tt
  2105.10161}}].

\bibitem{Witten:2021nzp}
E.~Witten, {\it {A Note On Complex Spacetime Metrics}},
  \href{http://xxx.lanl.gov/abs/2111.06514}{{\tt 2111.06514}}.

\bibitem{berry1972semiclassical}
M.~V. Berry and K.~Mount, {\it Semiclassical approximations in wave mechanics},
   {\em Reports on Progress in Physics} {\bf 35} (1972), no.~1 315.

\bibitem{heller1984bound}
E.~J. Heller, {\it Bound-state eigenfunctions of classically chaotic
  hamiltonian systems: scars of periodic orbits},  {\em Physical Review
  Letters} {\bf 53} (1984), no.~16 1515.

\bibitem{berry1985semiclassical}
M.~V. Berry, {\it Semiclassical theory of spectral rigidity},  {\em Proceedings
  of the Royal Society of London. A. Mathematical and Physical Sciences} {\bf
  400} (1985), no.~1819 229--251.

\bibitem{casati2006quantum}
G.~Casati and B.~Chirikov, {\em Quantum Chaos}.
\newblock Cambridge University Press, 2006.

\bibitem{gutzwiller2013chaos}
M.~C. Gutzwiller, {\em Chaos in classical and quantum mechanics}, vol.~1.
\newblock Springer Science \& Business Media, 2013.

\bibitem{heller2018semiclassical}
E.~J. Heller, {\em The semiclassical way to dynamics and spectroscopy}.
\newblock Princeton University Press, 2018.

\bibitem{simons2002mesoscopic}
B.~Simons and A.~Altland, {\it Mesoscopic physics},  in {\em Theoretical
  Physics at the End of the Twentieth Century}, pp.~451--566.
\newblock Springer, 2002.

\bibitem{Pollack:2020gfa}
J.~Pollack, M.~Rozali, J.~Sully, and D.~Wakeham, {\it {Eigenstate
  Thermalization and Disorder Averaging in Gravity}},  {\em Phys. Rev. Lett.}
  {\bf 125} (2020), no.~2 021601,
  [\href{http://xxx.lanl.gov/abs/2002.02971}{{\tt 2002.02971}}].

\bibitem{Dodelson:2022eiz}
M.~Dodelson and A.~Zhiboedov, {\it {Gravitational orbits, double-twist mirage,
  and many-body scars}},  \href{http://xxx.lanl.gov/abs/2204.09749}{{\tt
  2204.09749}}.

\bibitem{Collier:2022emf}
S.~Collier and E.~Perlmutter, {\it {Harnessing S-Duality in $\mathcal{N}=4$ SYM
  \& Supergravity as $SL(2,\mathbb{Z})$-Averaged Strings}},
  \href{http://xxx.lanl.gov/abs/2201.05093}{{\tt 2201.05093}}.

\bibitem{Gromov:2009zb}
N.~Gromov, V.~Kazakov, and P.~Vieira, {\it {Exact Spectrum of Planar ${\cal
  N}=4$ Supersymmetric Yang-Mills Theory: Konishi Dimension at Any Coupling}},
  {\em Phys. Rev. Lett.} {\bf 104} (2010) 211601,
  [\href{http://xxx.lanl.gov/abs/0906.4240}{{\tt 0906.4240}}].

\bibitem{Erickson:2000af}
J.~K. Erickson, G.~W. Semenoff, and K.~Zarembo, {\it {Wilson loops in N=4
  supersymmetric Yang-Mills theory}},  {\em Nucl. Phys. B} {\bf 582} (2000)
  155--175, [\href{http://xxx.lanl.gov/abs/hep-th/0003055}{{\tt
  hep-th/0003055}}].

\bibitem{Jensen:2018rxu}
K.~Jensen, A.~O'Bannon, B.~Robinson, and R.~Rodgers, {\it {From the Weyl
  Anomaly to Entropy of Two-Dimensional Boundaries and Defects}},  {\em Phys.
  Rev. Lett.} {\bf 122} (2019), no.~24 241602,
  [\href{http://xxx.lanl.gov/abs/1812.08745}{{\tt 1812.08745}}].

\bibitem{Chalabi:2020iie}
A.~Chalabi, A.~O'Bannon, B.~Robinson, and J.~Sisti, {\it {Central charges of 2d
  superconformal defects}},  {\em JHEP} {\bf 05} (2020) 095,
  [\href{http://xxx.lanl.gov/abs/2003.02857}{{\tt 2003.02857}}].

\bibitem{Iliesiu:2021are}
L.~V. Iliesiu, M.~Kologlu, and G.~J. Turiaci, {\it {Supersymmetric indices
  factorize}},  \href{http://xxx.lanl.gov/abs/2107.09062}{{\tt 2107.09062}}.

\bibitem{marolf2021ads}
D.~Marolf and J.~E. Santos, {\it {AdS Euclidean wormholes}},
  \href{http://xxx.lanl.gov/abs/2101.08875}{{\tt 2101.08875}}.

\bibitem{Arkani-Hamed:2007cpn}
N.~Arkani-Hamed, J.~Orgera, and J.~Polchinski, {\it {Euclidean wormholes in
  string theory}},  {\em JHEP} {\bf 12} (2007) 018,
  [\href{http://xxx.lanl.gov/abs/0705.2768}{{\tt 0705.2768}}].

\bibitem{affleck1984dynamical}
I.~Affleck, M.~Dine, and N.~Seiberg, {\it Dynamical supersymmetry breaking in
  supersymmetric qcd},  {\em Nuclear Physics B} {\bf 241} (1984), no.~2
  493--534.

\bibitem{Mahajan2021wormholes}
R.~Mahajan, D.~Marolf, and J.~E. Santos, {\it The double cone geometry is
  stable to brane nucleation},  \href{http://xxx.lanl.gov/abs/2104.00022}{{\tt
  2104.00022}}.

\bibitem{Brown:1992bq}
J.~D. Brown and J.~W. York, Jr., {\it {The Microcanonical functional integral.
  1. The Gravitational field}},  {\em Phys. Rev. D} {\bf 47} (1993) 1420--1431,
  [\href{http://xxx.lanl.gov/abs/gr-qc/9209014}{{\tt gr-qc/9209014}}].

\bibitem{Brown:1993ke}
J.~D. Brown and J.~W. York, Jr., {\it {Microcanonical action and the entropy of
  a rotating black hole}},  {\em Math. Phys. Stud.} {\bf 15} (1994) 23--34,
  [\href{http://xxx.lanl.gov/abs/gr-qc/9303012}{{\tt gr-qc/9303012}}].

\bibitem{Marolf:2018ldl}
D.~Marolf, {\it {Microcanonical Path Integrals and the Holography of small
  Black Hole Interiors}},  {\em JHEP} {\bf 09} (2018) 114,
  [\href{http://xxx.lanl.gov/abs/1808.00394}{{\tt 1808.00394}}].

\bibitem{Marolf:2022jra}
D.~Marolf and J.~E. Santos, {\it {Stability of the microcanonical ensemble in
  Euclidean Quantum Gravity}},  \href{http://xxx.lanl.gov/abs/2202.12360}{{\tt
  2202.12360}}.

\bibitem{balasubramanian1999stress}
V.~Balasubramanian and P.~Kraus, {\it {A stress tensor for anti-de Sitter
  gravity}},  {\em Communications in Mathematical Physics} {\bf 208} (1999),
  no.~2 413--428, [\href{http://xxx.lanl.gov/abs/hep-th/9902121}{{\tt
  hep-th/9902121}}].

\bibitem{prange1997spectral}
R.~Prange, {\it The spectral form factor is not self-averaging},  {\em Phys.
  Rev. Lett.} {\bf 78} (1997), no.~12 2280,
  [\href{http://xxx.lanl.gov/abs/chao-dyn/9606010}{{\tt chao-dyn/9606010}}].

\bibitem{Cotler:2016fpe}
J.~S. Cotler, G.~Gur-Ari, M.~Hanada, J.~Polchinski, P.~Saad, S.~H. Shenker,
  D.~Stanford, A.~Streicher, and M.~Tezuka, {\it {Black Holes and Random
  Matrices}},  {\em JHEP} {\bf 05} (2017) 118,
  [\href{http://xxx.lanl.gov/abs/1611.04650}{{\tt 1611.04650}}].

\bibitem{Cotler:2017jue}
J.~Cotler, N.~Hunter-Jones, J.~Liu, and B.~Yoshida, {\it {Chaos, Complexity,
  and Random Matrices}},  {\em JHEP} {\bf 11} (2017) 048,
  [\href{http://xxx.lanl.gov/abs/1706.05400}{{\tt 1706.05400}}].

\bibitem{dyson1962statisticalone}
F.~J. Dyson, {\it {Statistical theory of the energy levels of complex systems.
  I}},  {\em Journal of Mathematical Physics} {\bf 3} (1962), no.~1 140--156.

\bibitem{dyson1962statisticaltwo}
F.~J. Dyson, {\it {Statistical theory of the energy levels of complex systems.
  II}},  {\em Journal of Mathematical Physics} {\bf 3} (1962), no.~1 157--165.

\bibitem{dyson1962statisticalthree}
F.~J. Dyson, {\it {Statistical theory of the energy levels of complex systems.
  III}},  {\em Journal of Mathematical Physics} {\bf 3} (1962), no.~1 166--175.

\bibitem{mehta2004random}
M.~L. Mehta, {\it Random matrices},  {\em Pure and Applied Mathematics
  (Amsterdam). Elsevier/Academic Press, Amsterdam} (2004) 9.

\bibitem{haake2013quantum}
F.~Haake, {\em Quantum Signatures of Chaos}, vol.~54.
\newblock Springer Science \& Business Media, 2013.

\bibitem{Ambjorn:1990wg}
J.~Ambjorn and Y.~M. Makeenko, {\it {Properties of Loop Equations for the
  Hermitean Matrix Model and for Two-dimensional Quantum Gravity}},  {\em Mod.
  Phys. Lett. A} {\bf 5} (1990) 1753--1764.

\bibitem{brezin1993universality}
E.~Br{\'e}zin and A.~Zee, {\it Universality of the correlations between
  eigenvalues of large random matrices},  {\em Nuclear Physics B} {\bf 402}
  (1993), no.~3 613--627.

\bibitem{efetov1983supersymmetry}
K.~Efetov, {\it Supersymmetry and theory of disordered metals},  {\em advances
  in Physics} {\bf 32} (1983), no.~1 53--127.

\bibitem{efetov1999supersymmetry}
K.~Efetov, {\em Supersymmetry in disorder and chaos}.
\newblock {Cambridge University Press}, 1999.

\bibitem{Altland:2020ccq}
A.~Altland and J.~Sonner, {\it {Late time physics of holographic quantum
  chaos}},  {\em SciPost Phys.} {\bf 11} (2021) 034,
  [\href{http://xxx.lanl.gov/abs/2008.02271}{{\tt 2008.02271}}].

\bibitem{Altland:2017eao}
A.~Altland and D.~Bagrets, {\it {Quantum ergodicity in the SYK model}},  {\em
  Nucl. Phys. B} {\bf 930} (2018) 45--68,
  [\href{http://xxx.lanl.gov/abs/1712.05073}{{\tt 1712.05073}}].

\bibitem{Altland:2021rqn}
A.~Altland, D.~Bagrets, P.~Nayak, J.~Sonner, and M.~Vielma, {\it {From operator
  statistics to wormholes}},  {\em Phys. Rev. Res.} {\bf 3} (2021), no.~3
  033259, [\href{http://xxx.lanl.gov/abs/2105.12129}{{\tt 2105.12129}}].

\bibitem{d1995universal}
J.~D'anna, E.~Br{\'e}zin, and A.~Zee, {\it {Universal spectral correlation
  between Hamiltonians with disorder II}},  {\em Nuclear Physics B} {\bf 443}
  (1995), no.~3 433--443.

\bibitem{wilkinson1999parametric}
M.~Wilkinson, {\it {Parametric Random Matrices: Static and Dynamic
  Applications}},  {\em Supersymmetry and Trace Formulae} (1999) 369--399.

\bibitem{Meltzer:2017rtf}
D.~Meltzer and E.~Perlmutter, {\it {Beyond $a = c$: gravitational couplings to
  matter and the stress tensor OPE}},  {\em JHEP} {\bf 07} (2018) 157,
  [\href{http://xxx.lanl.gov/abs/1712.04861}{{\tt 1712.04861}}].

\bibitem{Maldacena:2001kr}
J.~M. Maldacena, {\it {Eternal black holes in anti-de Sitter}},  {\em JHEP}
  {\bf 04} (2003) 021, [\href{http://xxx.lanl.gov/abs/hep-th/0106112}{{\tt
  hep-th/0106112}}].

\bibitem{Herzog:2002pc}
C.~P. Herzog and D.~T. Son, {\it {Schwinger-Keldysh propagators from AdS/CFT
  correspondence}},  {\em JHEP} {\bf 03} (2003) 046,
  [\href{http://xxx.lanl.gov/abs/hep-th/0212072}{{\tt hep-th/0212072}}].

\bibitem{Maldacena:1997de}
J.~M. Maldacena, A.~Strominger, and E.~Witten, {\it {Black hole entropy in M
  theory}},  {\em JHEP} {\bf 12} (1997) 002,
  [\href{http://xxx.lanl.gov/abs/hep-th/9711053}{{\tt hep-th/9711053}}].

\end{thebibliography}\endgroup
\bibliographystyle{JHEP}

\end{document}